# Self-ordered Mo-oxide Nanotube Arrays as Precursor for Aligned MoOx/$MoS_2$ Core-Shell Nanotubular Structures with a High Density of Reactive Sites

Bowen Jin, [a] Xuemei Zhou, [b] Li Huang, [a] Markus Licklederer, [b]

Min Yang *[a] and Patrik Schmuki *[b]

---

[a] Bowen Jin, Li Huang, and Prof. Dr. Min Yang, School of Chemical Engineering and Technology, Harbin Institute of Technology, Harbin, 150001, PR China. E-mail: yangmin@hit.edu.cn

[b] Xuemei Zhou, Markus Licklederer, and Prof. Dr. Patrik Schmuki, Department of Materials Science, Institute for Surface Science and Corrosion (LKO), University of Erlangen-Nuremberg, Martensstr. 7, 91058, Erlangen, Germany. E-mail: schmuki@ww.uni-erlangen.de

**Abstract:** In the present work we demonstrate the self-organized formation of anodic Mo-oxide nanotube arrays grown on a Mo sheet under suitable electrochemical conditions in glycerol/$NH_4F$ electrolytes. The resulting amorphous tubes can be crystallized by annealing to $MoO_2$ or $MoO_3$. The tube walls then can be further sulfurized fully or partially to Mo-sulfide to form well-ordered arrays of vertically aligned MoOx/$MoS_2$ nanotubes. Under optimized conditions, defined $MoS_2$ sheets form on the oxide walls in a layer by layer low angle zig-zag arrangement that provide a high density of reactive stacking faults. These core-shell nanotube arrays, consisting of tubes with a conductive suboxide core and a functional high defect density $MoS_2$ coating, are highly promising for applications such as electrocatalysis (hydrogen evolution) or ion insertion devices.

Molybdenum compounds – particularly oxides and sulfides [1] – have for decades received attention due to their unique chemical and physical properties and the resulting high potential not only in classic lubrication[2] but also in photochromic/ electrochromic devices[3], photocatalysts[4], high energy density super-capacitors[5], Li-ion batteries[6], and most recently as catalyst for electrochemical hydrogen generation [1b, 7]. For many of these applications, the use of nanoscale geometries is highly advantageous due to the high specific surface area and short carrier and ion diffusion pathways that can be established. [8] Among the various MoX materials, particularly $MoS_2$ is intensively investigated. This is to a large extent due to the weak van der Waals force between the stacked S-Mo-S units giving it a graphene-like layer structure [9]. This is not only the origin of outstanding lubrication properties but also the key to the insertion and extraction of small foreign ions into the free space of the S-Mo-S layers [10].

In insertion devices the availability of the layer edges plays a crucial role to provide access for ions to enter the galleys between the sheets. Layer edges also are regarded crucial for the remarkable electrocatalytic properties observed for $MoS_2$ for the promotion of the hydrogen evolution reaction (HER). The synthesis of defined $MoS_2$ nanostructures was pioneered more than 20 years ago by the finding of Tenne et al[11] on the thermal synthesis of “onion-shell” $MoS_2$ structures. Meanwhile various other synthesis paths for $MoS_2$-based nano-morphologies have been explored – these are mainly based on either gas-phase reactions [9-12], electrodeposition [13], or wet-chemical template-assisted approaches [14]. In electro- or photoelectrochemical applications particularly 1D nanostructures (such as tubes or belts) have been shown to be highly effective in enhancing the performance of e.g. electrocatalytic or ion-insertion devices.

In contrast, $MoO_x$ and its 1D structures are generally found to be less effective to provide catalytic functionality and show generally lower ion insertion capacities, but the oxide and even more oxide/suboxide structures can provide a very high electron-conductivity [15].

Thus a particularly elegant combination of oxide and sulfide is the formation of core-shell structures (preferably vertically aligned to an electrode back contact) where the conductive oxide core ($MoO_3$/$MoO_2$) provides a charge transport pathway, and the $MoS_2$ shell enables above mentioned functionalities[16]. A general synthesis strategy is to form oxide structures that then can be converted to $MoO_x$/$MoS_2$ structures. Up to date, the synthesis of defined 1D $MoO_3$/$MoS_2$ co-nanostructures has mainly been achieved by hydrothermal approaches [17] or catalyzed transport reactions [18].

However, a most direct and facile approach to grow aligned 1D MoOx-nanotubular structures that are directly back-contacted would be self-ordering electrochemical anodization (SOA) of a metallic Mo substrate. SOA, over the past ten years, has been successfully developed and used to obtain highly ordered oxide nanotube or nanopore arrays on Al, Ti, W, Ta, Nb, Zr, Co, Fe and a variety of alloys. [19] Nevertheless, in spite of the high scientific and technological driving force to achieve self-ordered oxide growth on Mo, there is, up to now, *no report* on the successful fabrication of self-ordered Mo-oxide nanoporous or nanotubular arrays. This can be ascribed to the fact that to achieve a self-ordered growth of an anodic oxide structure, one needs to be able to establish suitable electrochemical conditions – specifically, a defined equilibrium between anodic oxide formation and its dissolution[19b, 20] − here not only chemical but also mechanical properties of a specific oxide are important to achieve stable nanotube growth. However, Mo belongs to the class of metals where an optimized equilibrium is very hard to be experimentally established. This is particularly due to the high solubility of Mo (VI) in many electrolytes – the oxide in many cases can easily undergo oxidative dissolution [21].

In the present work, we establish experimental conditions that enable the anodic growth of one-dimensional ordered Mo-oxide nanotube arrays [Fig. 1]. These amorphous tubes can be converted to various crystalline oxide phases. Furthermore, we show that a partial conversion of these oxide tubes in $H_2S$ can yield a unique oxide/sulfide core shell structure where the conducting suboxide core is coated with conformal $MoS_2$ layers in a low angle, zig-zag configuration that provides a high stacking fault density [Fig. 1b]. These structures show very promising first results in applications, such as, as an electrocatalyst for HER and as an anode for Li-ion battery configuration.

To achieve self-organized growth of Mo oxide structures, we explored a wide range of electrochemical conditions (an overview is given in the SI, Table S1) that provided potentially the pre-requisites for anodic tube growth [19]. Overall, the parameter range leading to self-organization is remarkably sensitive to variations such as water content or anodization voltage (see SI). However, from these experiments, we identified fluoride-containing glycerol electrolytes as most promising, and finally established a parameter range where self-organized oxide growth can be successfully and reliably achieved (as described in detail in the SI). Fig. 1a gives an example of a self-ordered Mo-oxide layer grown at 35 V for 1 h in the optimized electrolyte ($NH_4F/H_2O$/glycerol). Under these conditions, the nanotubes have a length of approx. 1 µm and an inner diameter of ≈80 nm. The layers adhere very well to the substrate and exhibit long range order, i.e. coat uniformly the entire anodized surface.

By a variation of the anodization voltage and time, the tube diameter and length can be controlled over a certain range (Fig. S2b). XRD of the nanotube layers shows the as-grown tubes to be amorphous (Fig. 2d), and XPS in Fig. 2 suggests the tube composition of these as-formed samples to be close to a $MoO_3$ stoichiometry.

To convert the amorphous layers to a defined crystallinity, samples were annealed under various conditions (250 °C, 350 °C, 450 °C in air and vacuum, respectively). Fig. S4-S6 shows XRD and SEM data of samples after the annealing experiments. The results demonstrate that a variety of amorphous and crystalline tubes can be produced using different thermal treatments. In this work we selected for further investigations annealing in vacuum as these tubes, in follow-up experiments, turned out to be most suitable for an optimized conversion to a $MoO_x/MoS_2$ core/shell structure. The corresponding XRD spectra (Fig. S4) reveal that annealing up to 250 °C, the XRD pattern remains the same as for the as-prepared amorphous Mo-oxide layer; for higher temperature one finds the formation of a desired mixed stoichiometry of $MoO_2$ and $MoO_3$. This is not only evident in XRD (Fig. 2d) but also from XPS (Fig. 2a) where the Mo peak in XPS for samples after vacuum annealing exhibits a shoulder at lower binding energy, which shows that some of the Mo remains reduced after the vacuum treatment[22] (as described in more detail in the SI). Most defined and stable crystalline conditions are provided by vacuum annealing at 350 °C. Under these conditions not only crystallinity but also the morphology of the nanotubes is maintained (optical and SEM images in Fig. 1b, and XRD patterns in Fig. 2) while higher temperatures lead to a thermal sintering of the structures (Fig. S5-S6).

These oxide tubes provide the basis for the formation of $MoO_x/MoS_2$ core-shell structures − this by a partial sulfurization as schematically shown in Fig. 1b. Various annealing treatments in $H_2S$ at different temperatures and durations were examined and the nanotubes then characterized in view of morphology, structure and composition (as described in more detail in the SI). The effect of the $H_2S$ treatment on the oxide can be well followed by XRD and XPS. Fig. 2 and S7 give examples for a treatment of the tube layers in $H_2S$ at 500°C for various exposure times. With increasing sulfurization time XRD and XPS spectra show a gradual conversion from oxide to sulfide. From XRD for sulfurization for more than 2 min, a clear peak at $2\theta = 14.3\,^{o}$ assigned to $MoS_2$ is observed. With prolonged treatment (60 min), there is an increase of the peak intensity of $MoS_2$, while the $MoO_3$ peaks disappear, and gradually the formation of $MoO_2$ peaks at 26.0° can be observed. This shows that with increasing $H_2S$ treatment time, besides sulfide formation, also a partial reduction of $MoO_3$ to $MoO_2$ takes place.

Fig. 2b shows XPS spectra for a sample treated for 2 min in $H_2S$. The $S_{2p}$ region indicates that at least two chemical states of S overlap. The doublet at 165.0 eV and 163.9 eV is assigned to S in a Mo-O-S configuration, the doublet of 163.8 eV and 162.5 eV to Mo-S bands, respectively [23]. After annealing in $H_2S$ for 4 min and 10 min the latter regime increases [24], consistent with the formation of additional $MoS_2$ with extended sulfurization time, the peak intensity of $O_{1s}$ decreases gradually – this in line with above XRD data and EDX measurements (Table S2).

SEM images taken after sulfurization (Fig. 1b and Fig. S7) show that layer thickness and tube diameter are maintained after the $H_2S$ treatment, however a slight roughening of the tube walls can be observed.

TEM for samples exposed to $H_2S$ for extended times (e.g. 30 min in Fig. 1b) show a full conversion of the oxide walls to $MoS_2$. More interesting are samples treated for some few minutes. For 10 s to 2 min $H_2S$ treatment times nanotube walls with core-shell structures can be produced. The TEM image (Fig. 1b, lower and Fig. S8) shows a tube wall of a sample treated for 2 min where four distinct layers of $MoS_2$ have been formed on the outer tube  wall. The layers are present conformally over the tube with layers grown with an offset of 10°- 15° to the tube normal leading to a zig-zag arrangement of the $MoS_2$-sheets, and a periodic occurrence of stacking faults in the $MoS_2$ layers, repeating approx. every 6 nm. XPS confirms the presence of $MoS_2$ (Fig. 2a) with $Mo^{4+}$ located at a lower binding energy (232.8 eV and 229.6 eV) and a sole peak at 227.4 eV corresponding to $S^{2-}$ [25]. Moreover, $Mo^{6+}$ and $Mo^{5+}$ peaks are still visible in the spectra – this confirming  the core to still consist of a mixed phase of $MoO_3$ and $MoO_2$ (in line with XRD (Fig. 2d)).

To explore the potential value of the core-shell structures we tested the tubes for their performance as an electrocatalytic HER evolution catalyst and carried out some preliminary experiments regarding the potential use as an anode in a Li-insertion battery configuration. Hydrogen evolution reaction (HER) measurements were carried out as described in the SI. Fig. 3a shows polarization curves of the anodic $MoO_3$ annealed for various heat-treatment conditions (in vacuum and in $H_2S$) in comparison with a Pt sheet. The onset potentials of all self-ordered $MoOx/MoS_2$ nanotube arrays are at approx. -110 mV to - 160 mV. The pure $MoO_3$ layers show only a very low activity, whereas the sulfurized samples (from 30 s to 60 min) show decades higher current densities at comparably low overpotentials. Among these samples, clearly the samples after 2 min-sulfurization – that is the core-shell structure in Fig. 1b - show the highest HER activity with a current density of -16 mA/cm$^2$ at -300 mV vs. RHE (-10 mA/cm$^2$ at -260 mV vs. RHE).

Samples that were treated for longer times in $H_2S$ exhibit a significant decrease in the HER activity (Fig. 3a). These data including Tafel plots (Fig. 3b) and a comparison with literature data in the SI (Table S3) demonstrate the beneficial effect of the core-shell structure. To test if this due to desired combination of $MoS_2$ functionality with oxide conductivity we acquired solid state IV curves (Fig. 3c). Indeed we find the conductivity of the core-shell structure (30.91 Ω) to be very similar to a reduced $MoO_3$ tube layer (21.03 Ω) while the tube layers converted to $MoS_2$ for longer times show a significantly higher resistivity (471.71 Ω). This shows that the synergy between core and shell operates well. Moreover, previous reports have shown that the edges of the two-dimensional $MoS_2$ catalysts are active sites for the HER reactions, especially when these basal planes are oriented off-axis to the sample surface [1b, 7a]. Hence the presence of the high number of stacking faults in the $MoS_2$ layers enhances the number of active sites for HER reactions. However, usually disorder, such as stacking faults, are reported to be highly detrimental to the intrinsic conductivity of $MoS_2$. Therefore the underlying conductive $MoO_x$ core (with a short carrier transport paths in $MoS_2$) is particularly important. It is also noteworthy that, in this case, the outside shell of $MoS_2$ not only provides active sites for HER activity but also provides the electrode stability against corrosion, as seen in Fig. S9-S11, and in line with literature[16] (see also further experiments in the SI).

While recently other very promising sulfide or phosphide compounds have been reported [26], the key of the present work is that it shows the direct facile formation of a $MoO_x/MoS_2$ core shell structure (nanotube arrays) that is directly back contacted, and thus can directly be used as an electrode. Moreover, it is noteworthy that even in the present state of optimization (see also Fig. S12), a slightly better performance can be achieved than any previous reported core-shell $MoO_3/MoS_2$ nano-structure [16].

As mentioned, the here synthesized core-shell structures are also of interest for Li-insertion devices [6], as a key issue in $MoS_2$ based battery devices is to combine the high Li-intercalation capacity of $MoS_2$ with conductivity (commonly addressed by using $MoS_2$/graphitic carbon composites). In the supporting information (Fig. S13) we give some preliminary data for the use of our 1D $MoO_x/MoS_2$ core-shell structures as an anode for lithium intercalation. First data show an initial specific capacity for the core shell structure to be in the range of 620 to 846 mA h $g^{-1}$ at 0.1 mA/$cm^2$. This represents, in comparison with other literature data, a very promising result (Table S4). This aspect will be followed up in detail in further work.

In summary, we show for the first time the successful synthesis of anodic self-organized $MoO_3$ nanotube layers. These layers could be used in classical applications of Mo-oxides such as solid-state lithium batteries, gas sensors, and electrochromic materials. The oxide nanotubular layer, however, can be further converted by sulfurization to $MoO_x$/$MoS_2$ core-shell nanotube structures. Remarkable is the high density of regular stacking faults in the outer $MoS_2$ sheets providing a high number of exposed reactive sites. These $MoO_x$/$MoS_2$ core shell structures thus exhibit an effective synergy between conductivity and reactivity that is evident from a high hydrogen evolution electrocatalytic activity that furthermore provides a promising platform for wider electrochemical and photoelectrochemical applications.

**Acknowledgement**

The authors thank Nhat Truong Nguyen for the TEM measurements, Gihoon Cha for the battery tests, Dr. Harald Gunselmann and Dr. Ringel Lorenz for the $H_2S$ detection, and Imgon Hwang for the solid state IV measurements. This work is financially supported by the National Natural Science Foundation of China (No.21203043), China Postdoctoral Science Foundation (No.2012M520729 and No.2013T60361), ERC, DFG, DFG-FUNCOS and the DFG Cluster of Excellence EAM.

**Keywords:** anodization • molybdenum oxide • nanotube • $H_2$ evolution • core-shell

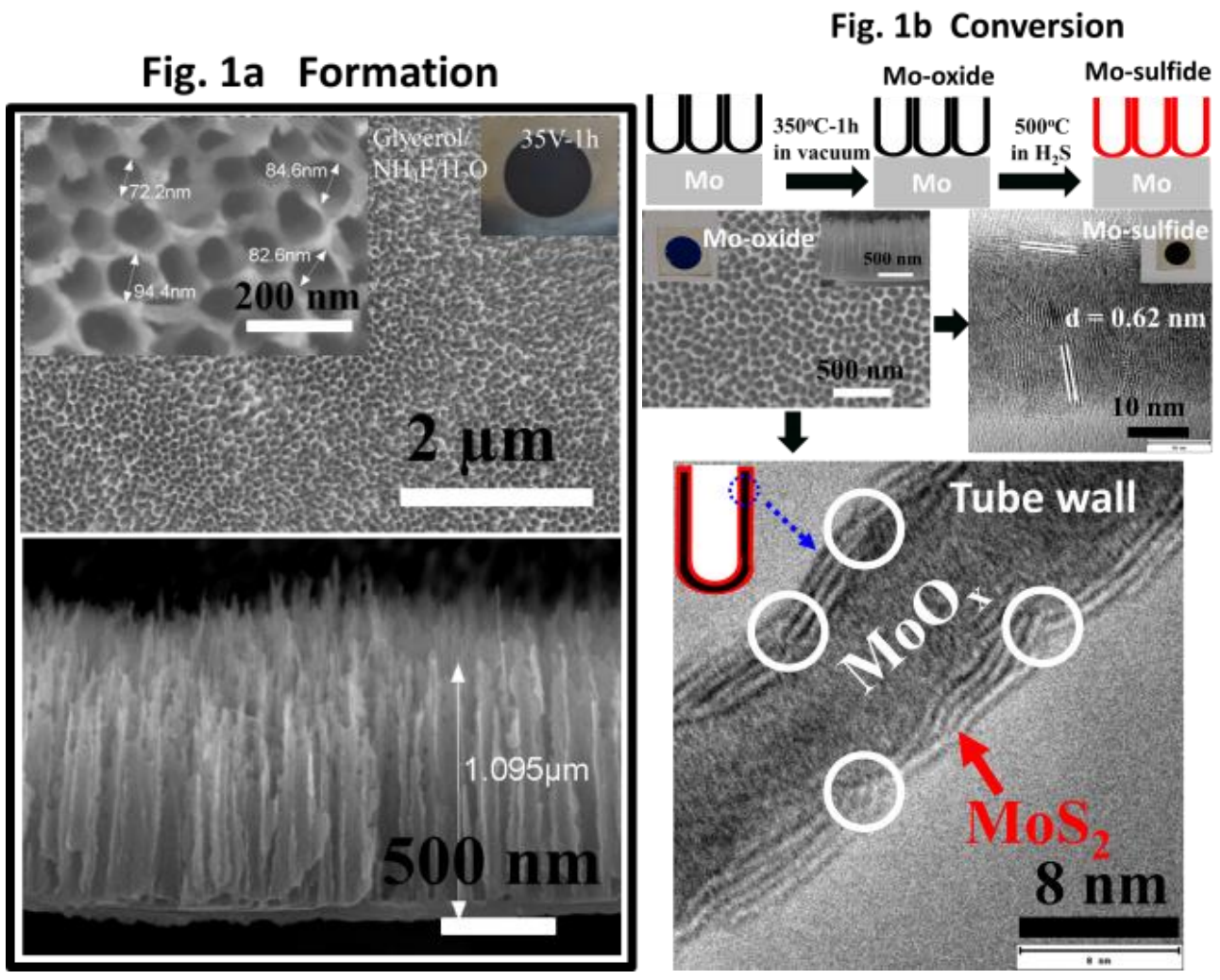

**Figure 1.** (a) SEM images for top view and cross-section (inset: optical image) of the Mo-oxide nanotubular structures anodized under optimized conditions (35 V for 1 h). (b) Thermal conversion from as-formed Mo-oxide to Mo-sulfide. Upper row: SEM images for Mo-oxide annealed at 350 °C for 1 h in vacuum; TEM image for fully conversion to $MoS_2$ (annealed in $H_2S$ at 500 °C for 30 min). Lower row: TEM image for partially conversion from Mo-oxide to MoOx/$MoS_2$ core-shell structure (annealed in $H_2S$ at 500 °C for 2 min). Inset illustrates the position where this TEM image taken from on the tube wall. The stacking faults of $MoS_2$ layers are marked in the white circles.

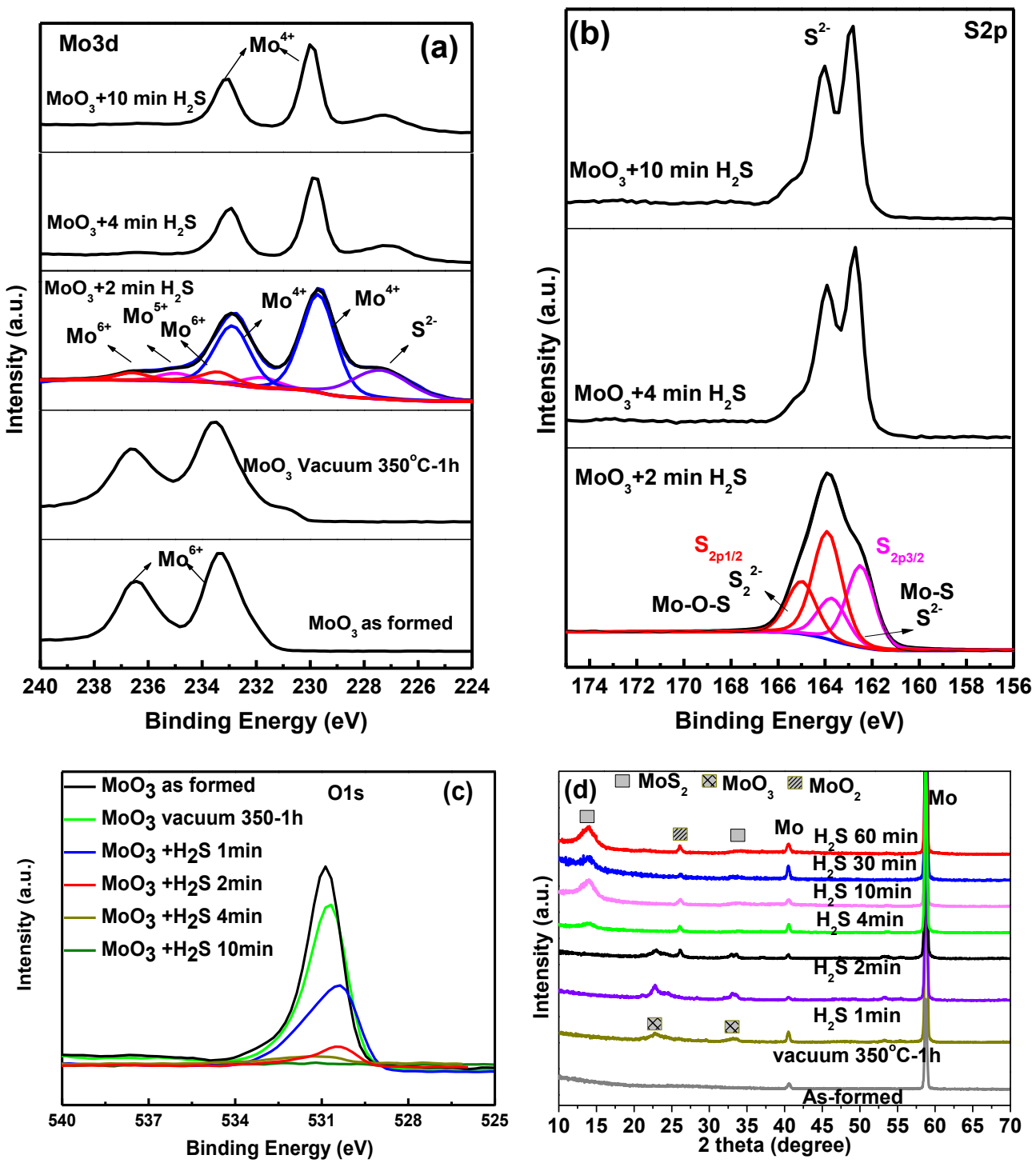

**Figure 2.** XPS spectra of $Mo_{3d}$ (a), $S_{2p}$ (b), and $O_{1s}$ (c) peaks of the samples before and after sulfurization in $H_2S$ for $MoO_3$ annealed in vacuum at 350 °C for 1h. (d) XRD patterns of $MoO_3$ nanotube layers after sulfurization.

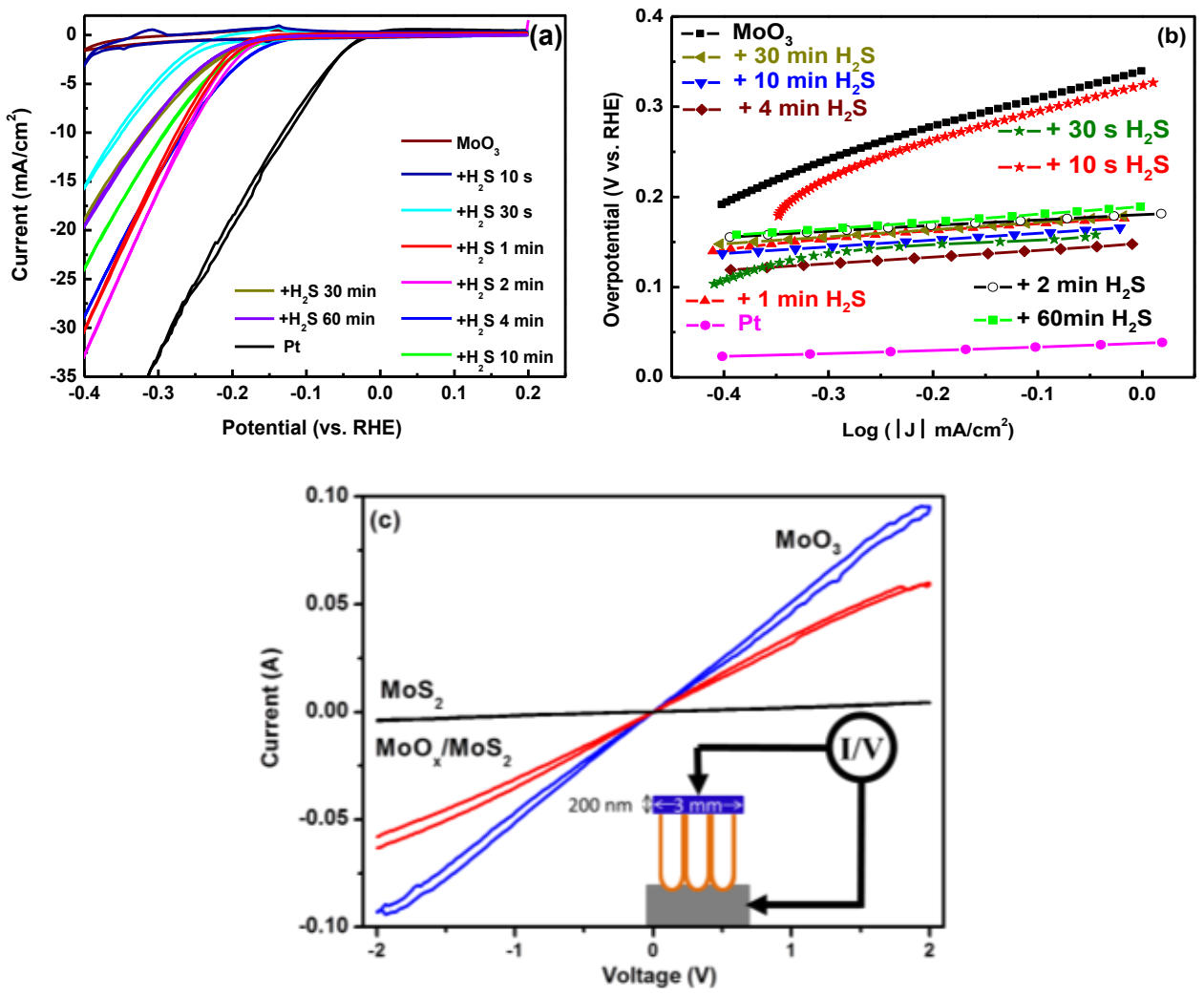

**Figure 3.** (a) Polarization curves for self-ordered $MoO_3$ nanoporous layers under various sulfurization time (10 s, 30 s, 1 min, 2 min, 4 min, 10 min, 30 min and 60 min, respectively), in comparison with a Pt sheet sample. Potential window is -0.4 to 0.2 V (vs. RHE) with scan rate of 5 mV/s. (b) Tafel analysis of the data presented in Fig. 3a. (c) Solid state I-V measurements for $MoO_3$ annealed at 350 °C for 1 h in vacuum, $MoOx/MoS_2$ and $MoS_2$ nanotube layers. Inset: illustrations for the Mo-oxide NT layers used for the IV measurements.

# Supporting Information

# Self-ordered Mo-oxide Nanotube Arrays as Precursor for Aligned $MoO_x/MoS_2$ Core-Shell Nanotubular Structures with a High Density of Reactive Sites

Bowen Jin [a], Xuemei Zhou [b], Li Huang [a], Markus Licklederer [b], Min Yang [a*] and Patrik Schmuki [b*]

[a] *School of Chemical Engineering and Technology, Harbin Institute of Technology, Harbin, 150001, PR China.*
[b] *Department of Materials Science, Institute for Surface Science and Corrosion (LKO), University of Erlangen-Nuremberg, Martensstrasse 7, 91058 Erlangen, Germany.*

Email: yangmin@hit.edu.cn, schmuki@ww.uni-erlangen.de

**Experimental Section**

One dimensional self-ordered molybdenum oxides were grown on Mo foils with 0.1 mm thickness and 99.6% purity. Before anodization, the foils were degreased by ultrasonication in acetone and ethanol successively, rinsed with deionized (DI) water, and dried in a nitrogen stream. Electrochemical anodization was carried out in a two-electrode system using a Pt sheet as a counter electrode. For the best anodization conditions, the substrates were polarized in glycerol-water electrolyte (volume ratio 9:1) containing 0.4 M $NH_4F$ at 35 V for 1 h to achieve an approx.1 μm thick nanoporous film. After anodization the resulting layers were kept in the electrolyte for 3-5 min, then immersed into ethanol to remove the rest of electrolyte, and dried in nitrogen.

To evaluate optimized crystallization by heat-treatment, as-formed nanotubes were annealed at 250 $^oC$, 350 $^oC$, and 450 $^oC$ in vacuum and in air, respectively. In general, the tube structure collapses when annealed in air (see optical and SEM images in Fig. S5 and S6). Besides, cyclic voltammetry curves in Fig. S4 shows that the sample annealed at 350 °C in vacuum has the highest CV response. Due to these reasons, we transfer the as-formed layers to a fully crystalline oxide layer by annealing the samples in vacuum at 350 °C.

Sulfurization was carried out in a tube furnace (Heraeus ROK/A6/30, 220 V, 11.4 A, 1-50 Hz, 2.5 KW) with a thermal controller (EUROTHERM, ZEW 4150-4) at a $H_2S$ (Linde Purity 2.5) volume flow of 6 L/h at 500 °C.

For the morphological characterization, a field emission scanning electron microscope (FE-SEM, Supra™55, Zeiss) or a Hitachi FE-SEM S4800 equipped with an energy dispersive X-ray (EDX) analyzer was used. The length of the nanoporous array was directly obtained from SEM cross-sections. X-ray diffraction analyses (XRD, D8-Advance, Bruker) with graphite monochromized CuKα radiation (λ= 0.15406 nm) was carried out for detecting the crystal structure. The composition and the chemical state of the anodic layers were characterized using X-ray photoelectron spectroscopy (XPS, PHI 5600, US).

Solid current–voltage (I–V) measurements: The Mo-oxide layers were coated with Pt dots (through a shadow mask of) on the top by depositing a nominally 200 nm thick Pt (3.0 mm diameter dot) layer using a plasma sputter device (EM SCD500, Leica) operated at 16 mA at $10^{-2}$ mbar vacuum, using Ar as moderating gas. Solid-state conductivity measurements were carried out by a 2-point measurement setup which consisted of a USMCO micromanipulator and precision semiconductor parameter analyzer (4156C,Agilent technologies, Japan). All I-V curves were measured with 20 mV/s sweep rate in the voltage window from -2 V to 2 V. Samples were treated at 110°C for 20 min before measurements and all the measurements were performed in an argon glove box.

For the electrochemical measurements, self-ordered $MoO_3$ or $MoO_x/MoS_2$ arrays on the Mo foil were employed as the anode, a graphite rod as a counter electrode, and a calomel electrode as a reference electrode. $H_2$ evolution measurements were carried out in 0.5 M $H_2SO_4$ aqueous solution. The gas mixture from the head space was measured by a Agilent Technologies 7890A Gas Chromatograph coupled with a 5975C Mass Spectrometer.

We also detected the $H_2S$ in the mixture gas by CV measurements. First, 1.0 mL sample gas was injected into 0.5 mL 1M NaOH in 1.5mL vessel, close the top cap with a gas tight seal. Then, we shaked the vessel for 1 min and kept it. Secondly, we degassed a supporting electrolyte (10 mL 0.1 M NaOH) for 120 s and measure the blank solution. Then, we added 0.1 mL of above prepared sample solution into the supporting electrolyte. Measurements were carried out using a Metrohm 797 VA Computrace work station, using CV in the differential pulse mode. Enrichment at -0.5 V for 10 s, sweeping from -0.5 V to -0.9 V with HMDE. Voltage step time: 0.1 s; Voltage step: 0.006 V; Pulse time: 0.4 s; Pulse: 50 mV; Working electrode: Mercury electrode; Reference electrode: Ag/AgCl; Counter electrode: Pt.

The conversion of potential is based on following equation:

E (RHE) = E (SCE) + 0.2363 V +0.0592 pH @ 25°C

In 0.5M $H_2SO_4$, E (RHE) = E (SCE) + 0.2363 V

Lithium ion insertion battery tests were performed in an Argon filled glove box (MBraun) with a water and oxygen content below 5 ppm in an electrochemical cell, exposing 1 $cm^2$ of the sample surface, using a lithium sheet as the counter electrode and the sample used as anode. The electrolyte consisted of 0.1 M $LiClO_4$ in ethylene carbonate. The cyclic voltammetry (CV) as well as the galvanostatic charge-discharge measurements were recorded with a potentiostat/galvanostat type PGV 1A-OEM and Autolab by Metrohm. The potential window was set between 0.01 V and 3.5 V.

The mass of sample layers were measured by following two methods:

1. The weight of the sample layers were measured with a balance with 0.01mg resolution. After weighing, the nanotube layers were removed by ultra-sonication and cracking. The leftover of the substrates was measured again. Overall, the average mass for one piece of oxide layer is 0.20 ± 0.05 mg. This procedure was repeated 5 times.
2. As a second approach we estimated he weight of the tube layer from equations below:

$$m = \rho V = \rho (V_{bulk}-V_{pores})$$

$$V_{bulk} = \pi r^2 h$$

$$V_{pores} = \pi (d_i/2)^2 h N$$

$$N = \pi r^2 / (\pi (d_o/2)^2)$$

Where $\rho$ is the density of $MoO_3$, 4.69 g/$cm^3$. $V_{bulk}$ is the volume for the whole layer without pores, r is the radius of samples layer, 0.5 cm, h is the height of the porous layer, 1.1 µm. $V_{pores}$ is the volume of the pores. $d_i$ is the inner diameter, average 80 nm. N is the number of the pores, which is calculated according to the area ratio. $d_o$ is the outer diameter, approx. 140 nm (based on TEM). Overall, the mass of one sample layer is calculated to be 0.273 mg. Therefore, we obtain the estimated specific capacity for a comparison with reported results, as shown in Table S3.

Table S1 A summary of electrolyte compositions, anodization conditions and resulting morphology of the obtained $MoO_3$ layers.

| Electrolyte | | | Voltage (V) | Film description |
|---|---|---|---|---|
| $NH_4F$ | Solvent (vol%) | $H_2O$ (vol%) | | |
| 0.4 M | Glycerol 100% | 0% | 30 V | A precipitate film which is completely dissolved by washing in ethanol. No self-ordered oxide structures, only some dimples left on Mo substrate. |
| 0.4 M | Glycerol 100% | 0% | 40 V | |
| 0.4 M | Glycerol 100% | 0% | 50 V | |
| 0.4 M | Glycerol 100% | 0% | 60 V | |
| 0.4 M | EG[a] 100% | 0% | 15 V | A yellow film, which is mostly dissolved when taken out from the electrolyte. Remainder shows no tubular or porous structure. |
| 0.4 M | EG 100% | 0% | 20 V | A yellow film, but completely dissolved by washing in ethanol. No self-ordered oxide structures, only some dimples left on Mo substrate. |
| 0.4 M | EG 100% | 0% | 30 V | |
| 0.4 M | EG 100% | 0% | 60 V | |
| 0.4 M | Glycerol 95% | 5% | 20 V | A blue film without nanotubular or nanoporous morphology. |
| 0.4 M | Glycerol 95% | 5% | 30 V | A blue film without nanotubular or nanoporous morphology. |
| 0.4 M | Glycerol 95% | 5% | 40 V | A blue film but completely dissolved in ethanol during cleaning (Fig. S1b). (Some blue particles left.) |
| 0.4 M | Glycerol 95% | 5% | 50 V | |
| 0.4 M | Glycerol 90% | 10% | 20 V | A film with uniform, regular nanotubes. (Fig. S1c, S2c) (short) |
| 0.4 M | Glycerol 90% | 10% | 50 V | A film with uniform, regular nanotubes. (Fig. S2d) (top etched) |
| 0.4 M | Glycerol 85% | 15% | 20 V | A green film with irregular and limited length of nanopores. |
| 0.4 M | Glycerol 85% | 15% | 30 V | A blue film with irregular and limited length of nanopores. |
| 0.4 M | Glycerol 85% | 15% | 40 V | A blue film with limited length of nanopores, irregular and slightly defective surface. (Fig. S1d) |
| 0.4 M | Glycerol 85% | 15% | 50 V | A blue film with limited length of nanopores, some dimples, irregular and severely cracked surface. |

EG[a]: ethylene glycol

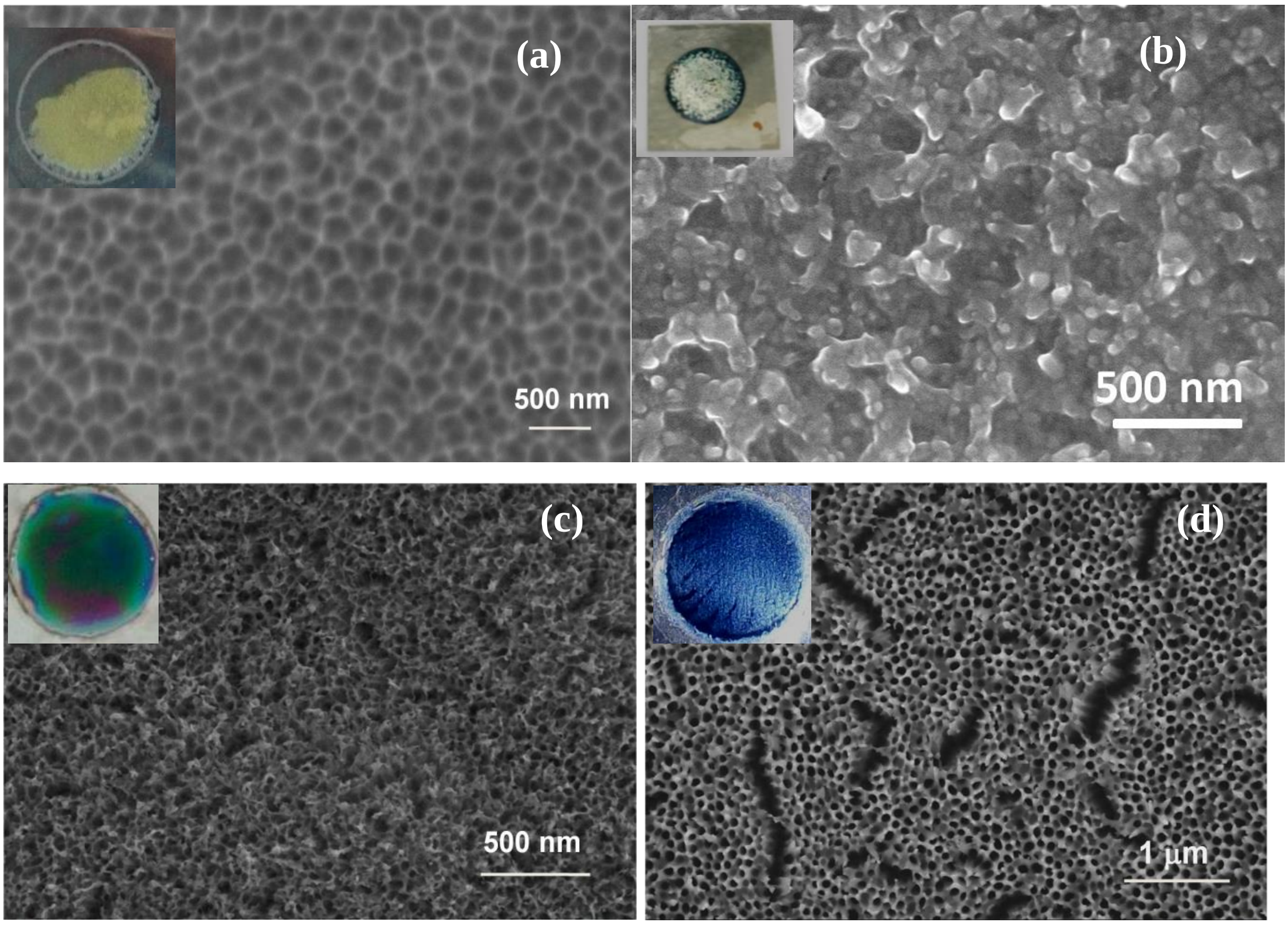

Fig.S1 SEM images (and optical photos) of anodically grown layers obtained under conditions of a) 100% glycerol containing 0.4 M $NH_4F$ at 30 V; (b) 95% glycerol containing 5% $H_2O$ and 0.4 M $NH_4F$ at 40 V; (c) 90% glycerol containing 10% $H_2O$ and 0.4 M $NH_4F$ at 20 V; (d) 85% glycerol containing 15% $H_2O$ and 0.4 M $NH_4F$ at 40 V.

The optical images of the surface in Fig. S1 and corresponding SEM images illustrate the effect of variation of the water content in the electrolyte. In general for water contents below 5%, only a yellow precipitate film (consist of nonstoichiometric $MoO_3$) without the formation of any organized structure is formed. This precipitate film dissolves quickly during washing with ethanol (see SEM in Fig. S1). For water contents >15% in the electrolyte, no uniform oxide layer can be formed. Keeping the water content in the range between 5% and 15% allows for a reproducible robust growth of nanotubes as in Fig. S1 c and d.

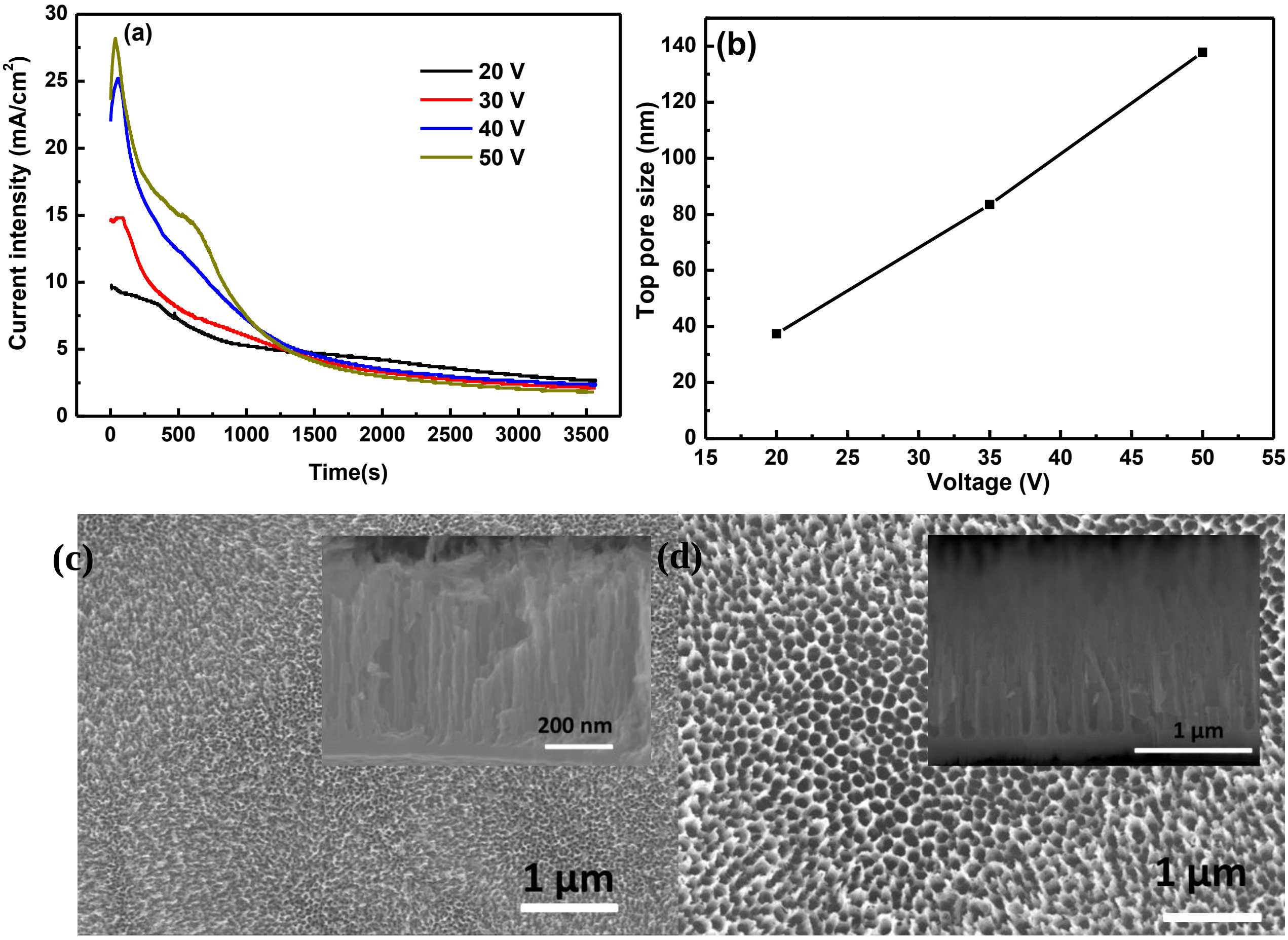

Fig. S2 (a) Current density transients of samples anodized at various applied voltages in 10%water/glycerol/$NH_4F$. In every case, the current initially drops further followed by leveling off to a steady-state value. The initial drop can be related to a growth of the Mo oxide layer. (b) Pore diameter as a function of the applied voltage. (c, d) SEM top view images (inset: cross-sections) of as-formed $MoO_3$ nanotubular layer at 20 V (c) and 50 V (d). This illustrates control over tube diameters.

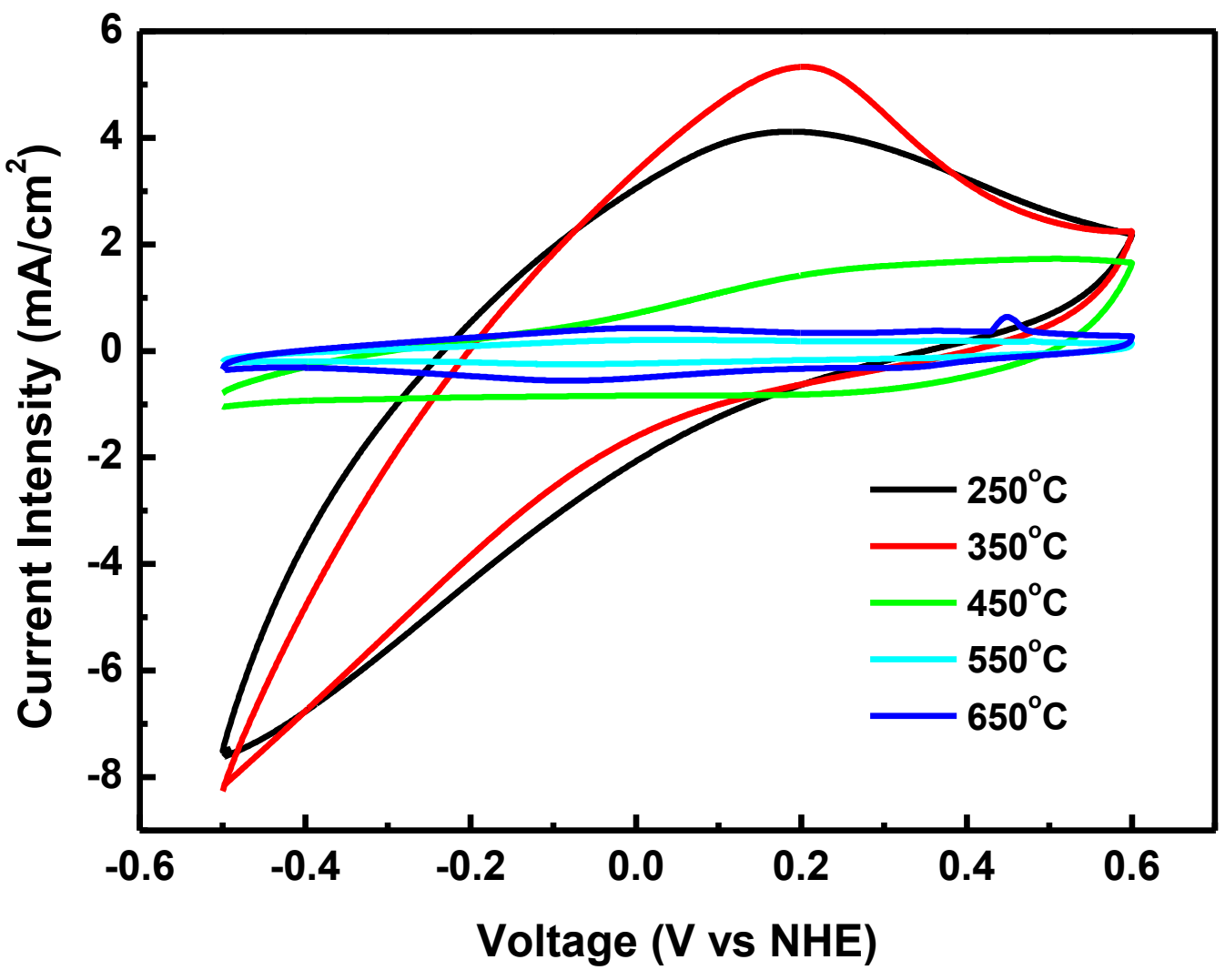

Fig.S3 Cyclic voltammograms of self-ordered anodic $MoO_3$ nanoporous layer after various annealing treatments in vacuum. The data were collected in 0.1M $HClO_4$ aqueous solution with rate of 100 mV/s. The best electrode performance is observed for 350 $^oC$ annealing, i.e. the nanotubular film of Fig.1 annealed to a $MoO_3$ crystalline structure (Fig. S4 XRD and Fig. 2a XPS).

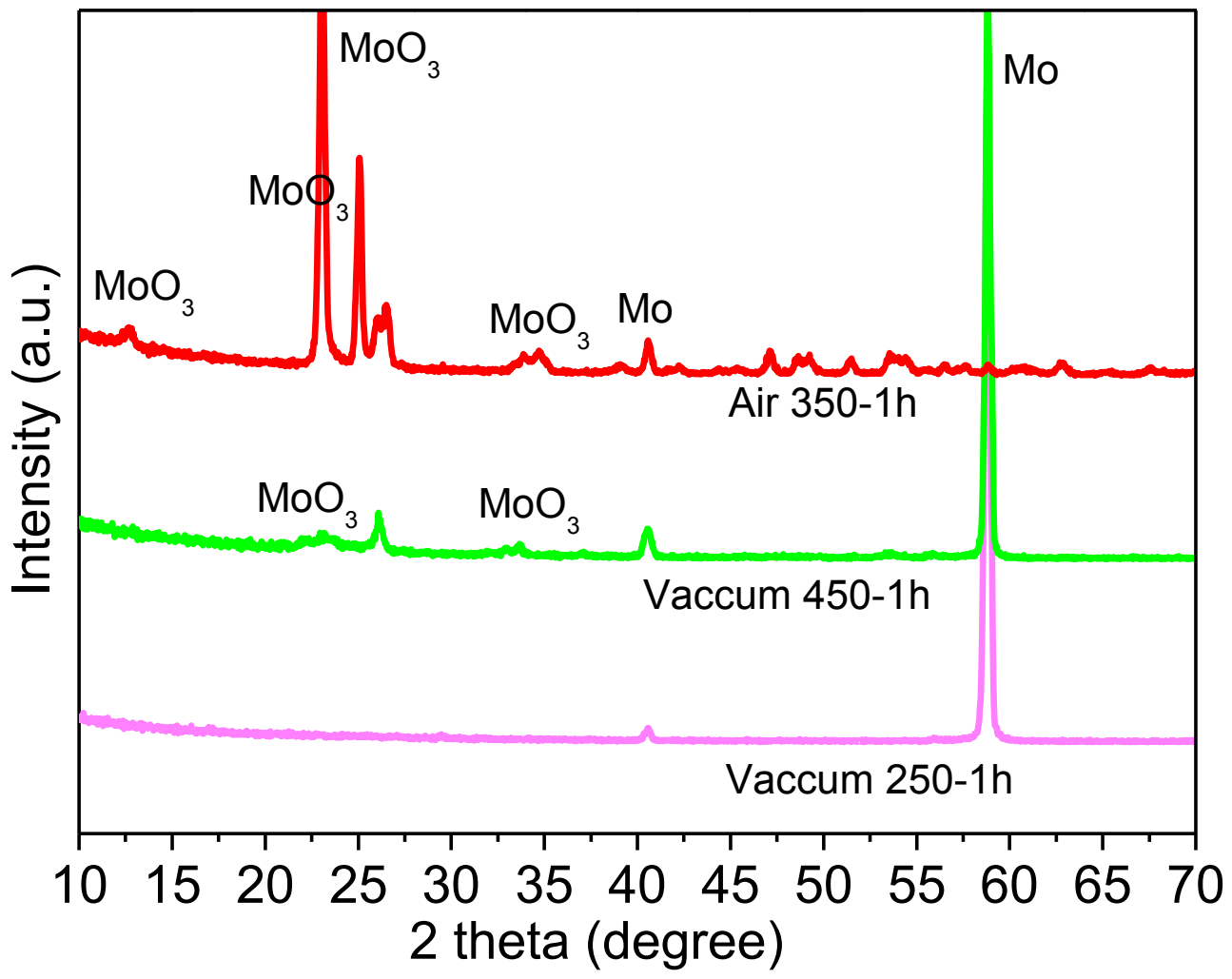

Fig.S4 XRD patterns for Mo-oxide layers annealed under different conditions.

The layers obtained at 35V with a 10% water content in the electrolyte is $MoO_3$, which shows in XPS two $Mo3d_{5/2}$ and $Mo3d_{3/2}$ peaks (Fig. 2) with respective binding energies around 233.4 eV and 236.6 eV, that can be assigned to $Mo^{6+}$ [S1]. XRD for this sample shows that the sample is amorphous with strong peaks at 40.6 ° and 58.8 ° that represent the Mo substrate (00-042-1120 Mo cubic). However, after annealing these layers in vacuum or in air, peaks at 2θ =22.8 ° and 33.1 ° appear, corresponding to crystallized $MoO_3$ (Fig. 2 and Fig. S4) [PDF card No. of 01-089-1554 $MoO_3$ Monoclinic].

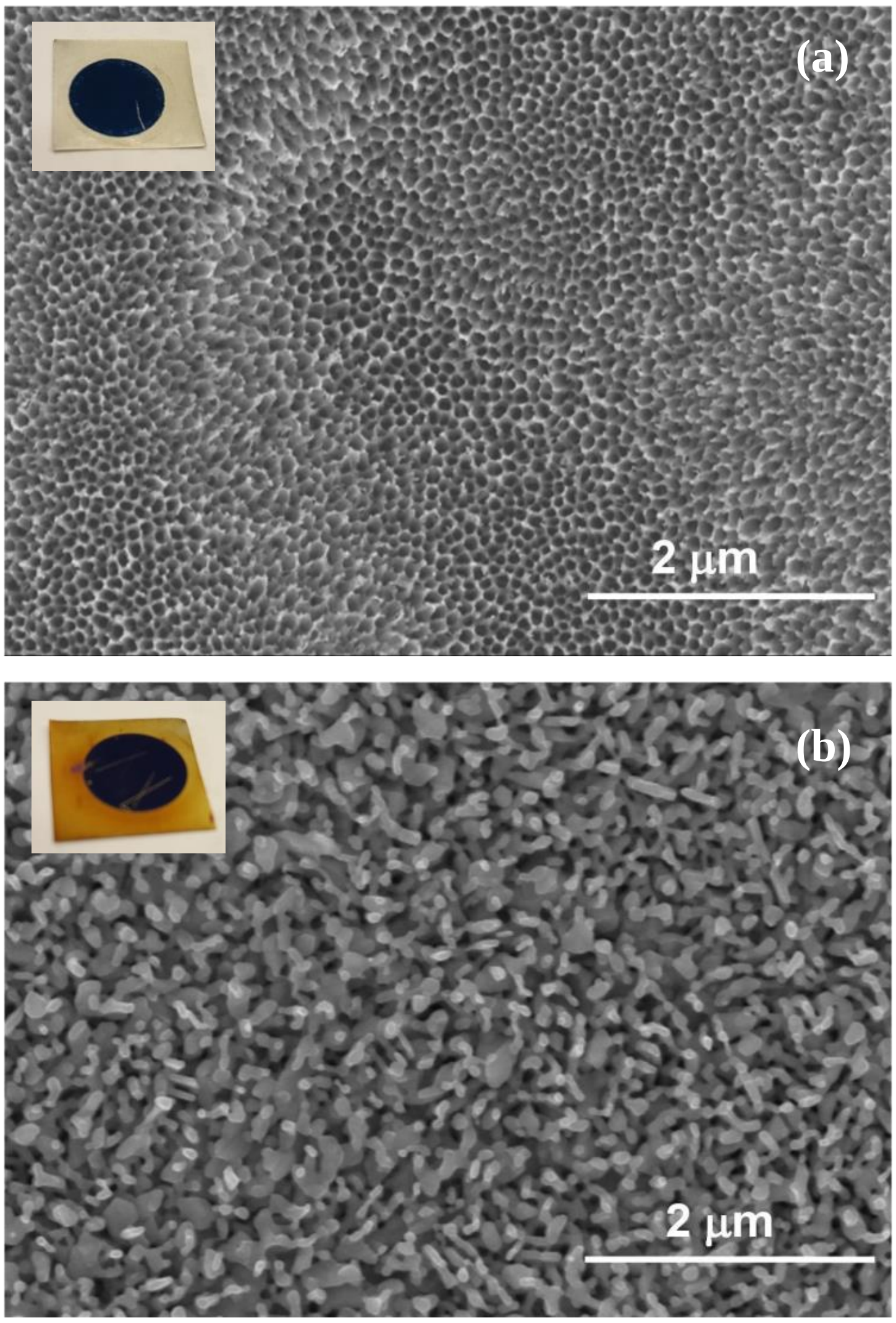

Fig. S5 SEM images for samples annealed in vacuum at 250 °C (a) and 450 °C (b), respectively. Annealing at 250 °C leads to intact tubular structure, however crystallization does not take place (Fig. S4). Annealing at 450 °C leads to structural collapse (sintering).

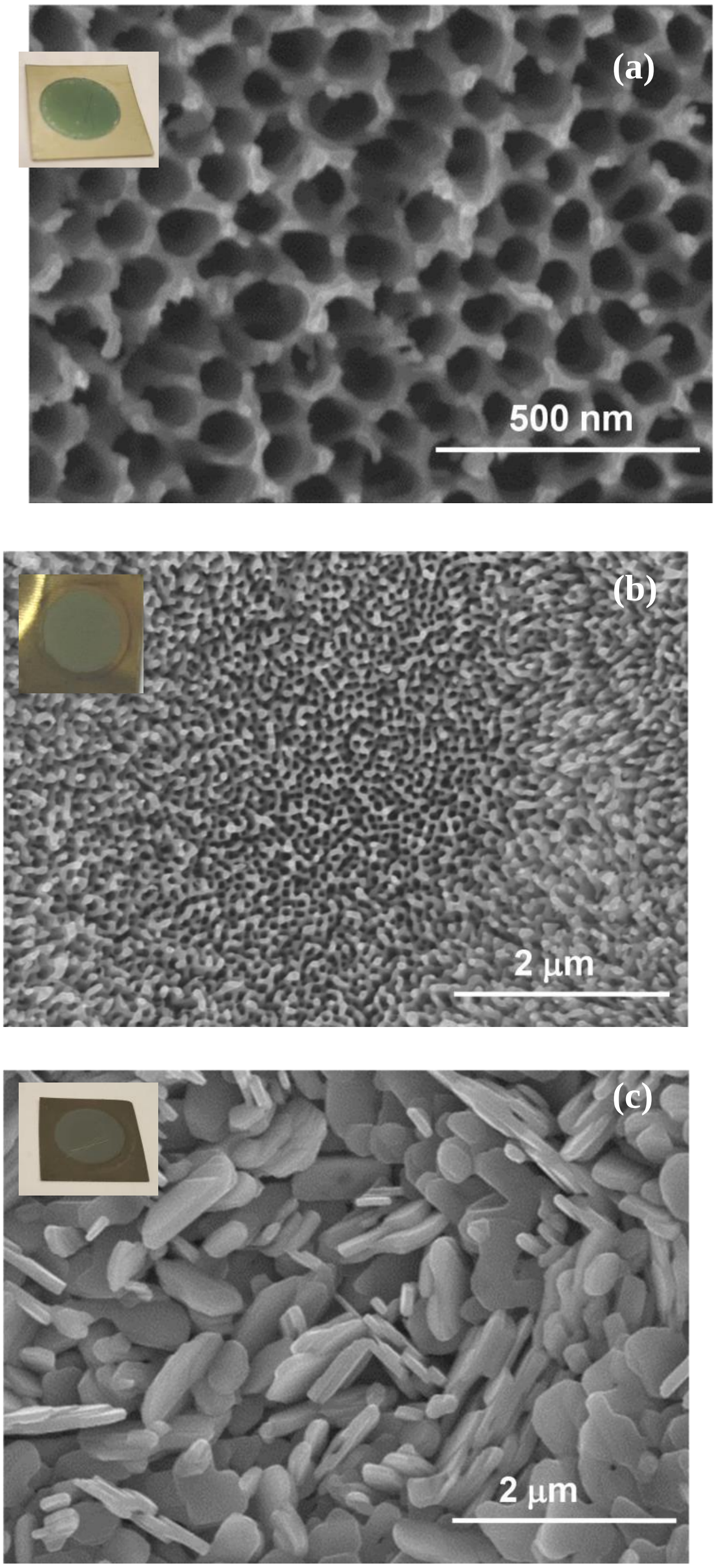

Fig.S6 SEM images of $MoO_3$ nanoporous layer annealed at 250 $^{o}C$ (a), 350 $^{o}C$ (b), and 450 $^{o}C$ (c) in air. (Please note that at 250 $^{\circ}C$ the layer is not crystallized, at 350 $^{\circ}C$ structural damage occurs, and at 450 $^{\circ}C$ the entire film is converted to a plate-like structure.)

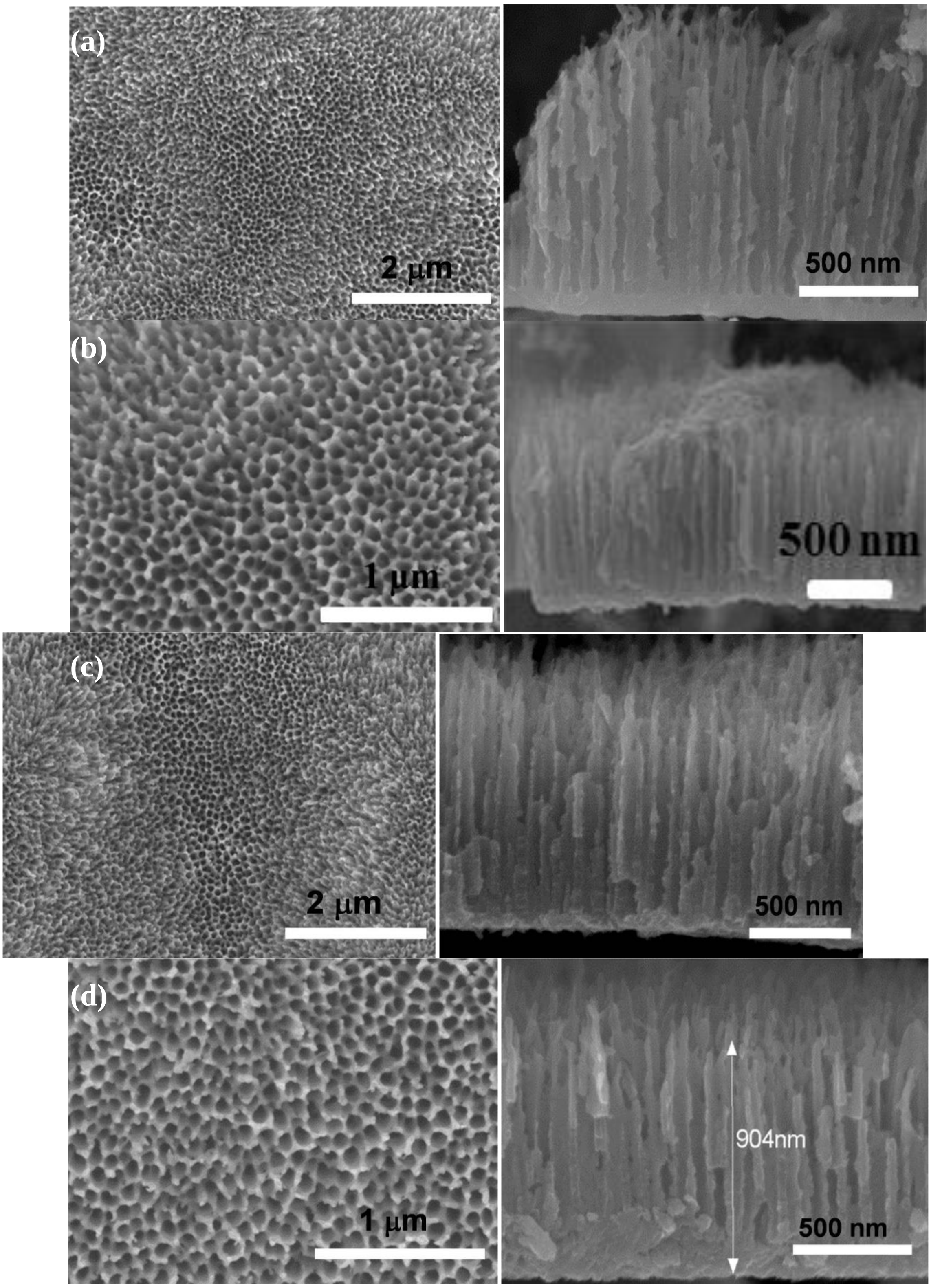

Fig.S7 SEM images of $MoO_3$ nanoporous layer annealed at 350 $^{o}C$ in vacuum and then $H_2S$ treated for 1min (a), 2 min (b), 4 min (c) and 30min (d).

We used annealed oxide tube layers and then sulfurized them in $H_2S$ gas at 500 °C for different times (10 s to 60 min) to explore the possibility to fabricate MoOx/$MoS_2$ core-shell structures. Fig. S7 shows the interface of the tube layer and substrate to be intact and flat with around 100 nm thickness. The sulfurization of the layer takes place from outmost side of the wall gradually into the inside wall and lower part of the tubes. A saturation of sulfurization is reached after a treatment time of 30 min, i.e. the XRD and performance are similar for samples treated for 30 min and 1h, respectively. The remaining $MoO_2$ is likely from remaining oxide at the interface of tube layer and substrate. The structure adheres well on the Mo back-contact. The S2p region from XPS (Fig. 2) can be split into 4 peaks, indicating that the $2p_{1/2}$ and $2p_{3/2}$ of at least two chemical states of S overlap. The peaks at 165.0 eV and 163.9 eV are assigned to bonds of Mo-O-S, while 163.8 eV and 162.5 eV are ascribed to Mo-S, respectively [S2].

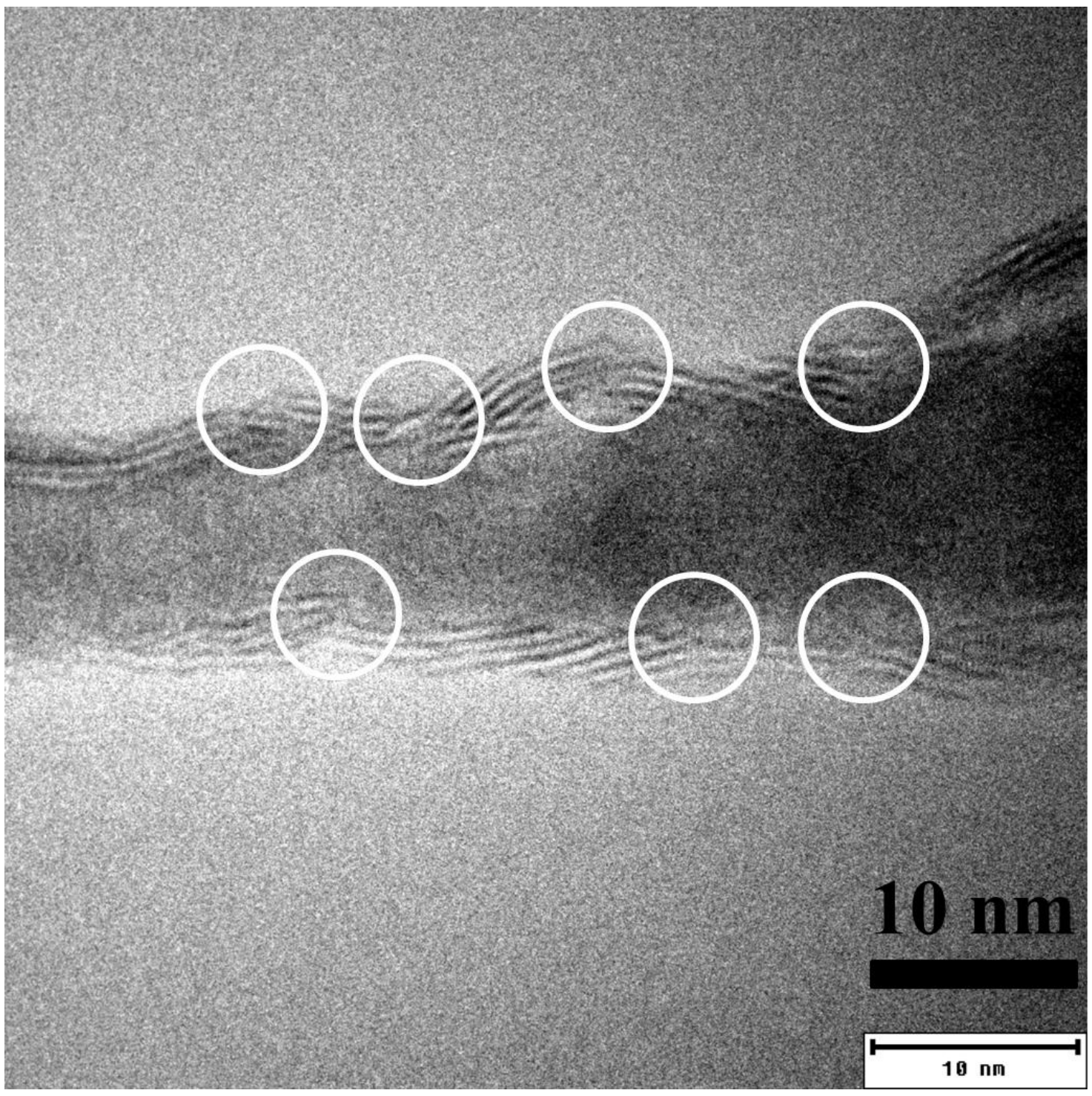

Fig. S8 TEM images of the wall of core-shell $MoO_x/MoS_2$ nanotube that is treated in $H_2S$ for 2min at 500 $^oC$.

Table S2 Composition of the structures as obtained from EDX spectra.

| Elements (at%) | **C** | **O** | **S** | **Mo** |
|---|---|---|---|---|
| $MoO_3$ | 10.3 | 46.01 | 0 | 43.68 |
| $MoO_3$ +$H_2S$ 1min | 11.74 | 42.92 | 7.05 | 38.28 |
| $MoO_3$ +$H_2S$ 2min | 8.50 | 35.96 | 12.91 | 42.64 |
| $MoO_3$ +$H_2S$ 4min | 5.81 | 5.36 | 52.81 | 36.03 |

In order to obtain further compositional information over the entire tube length we acquired EDX data. The results support a gradual formation of sulfide – this is deduced from a decrease of the oxygen peaks (direct evidence is hampered by partial overlap of sulfur and molybdenum peaks).

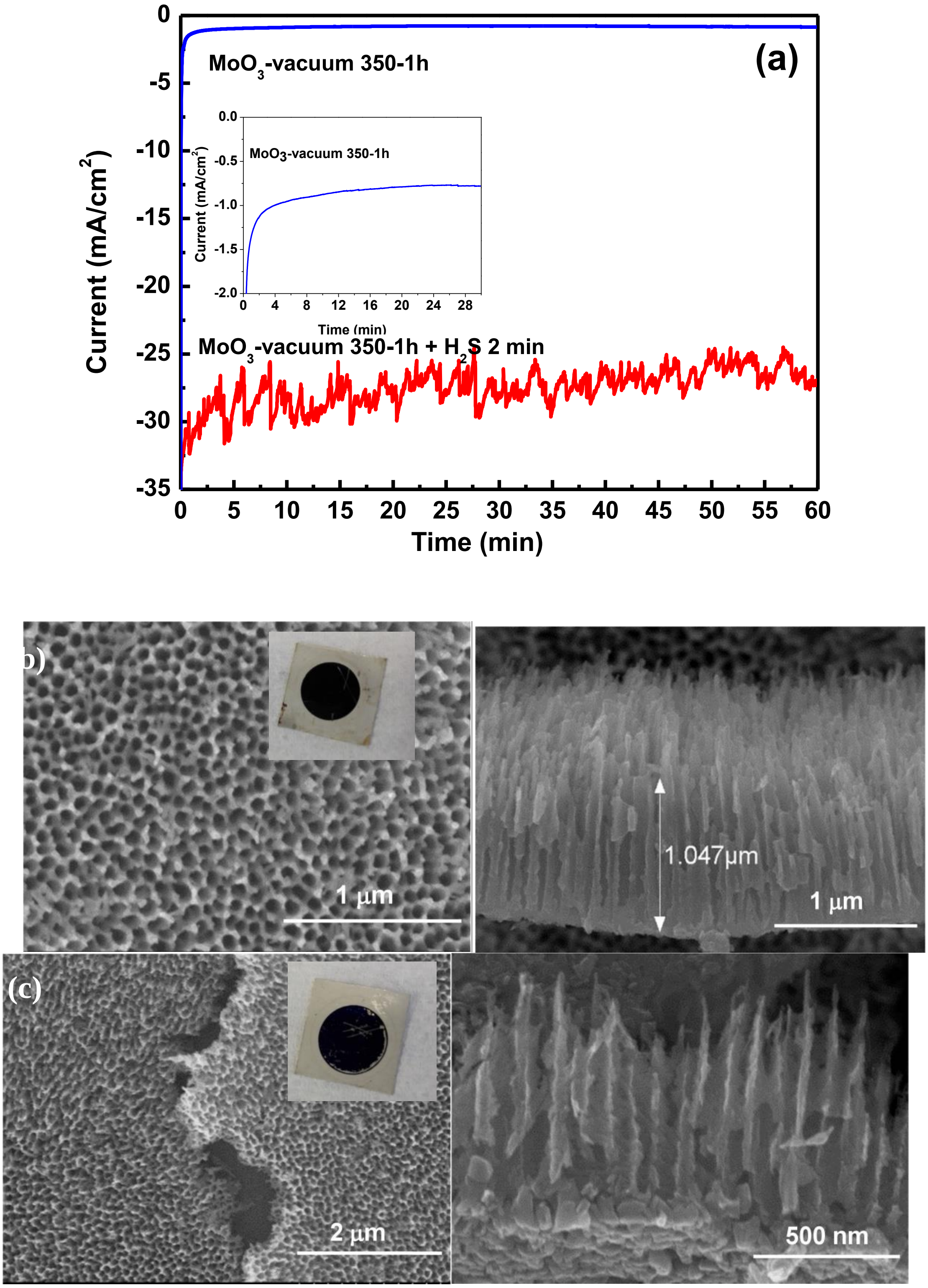

Fig.S9. (a) Current vs. time curves at -0.4 V (v.s. RHE) in 0.5 M $H_2SO_4$ for the core-shell structure (Fig. 1b lower row) and Mo-oxide (Fig. 1b upper row). (b) SEM image of $MoO_x/MoS_2$ nanoporous layer after $H_2$ evolution measurement. (c) SEM image of $MoO_3$ nanotubular layer after $H_2$ evolution measurement (inset: optical images for samples after stability measurement).

Table **S3.** Summary of Tafel slopes and onset potentials for $H_2$ evolution from own data (Fig. 3) and comparison with some literature data.

| Samples | Tafel slope (mV/dec) | Onset η (mV) | j at η = 300 mV ($mA/cm^2_{geometric}$) |
|---|---|---|---|
| $MoO_3$ | 318 | _ | 0.6 |
| $MoO_3$+10 s $H_2S$ | 419 | _ | 0.7 |
| $MoO_3$+30 s $H_2S$ | 152 | -200 | 5.2 |
| $MoO_3$+1 min $H_2S$ | 84 | -155 | 13.8 |
| $MoO_3$+2 min $H_2S$ | 63 | -147 | 15.9 |
| $MoO_3$+4 min $H_2S$ | 74 | -110 | 14.2 |
| $MoO_3$+10 min $H_2S$ | 74 | -134 | 10.8 |
| $MoO_3$+30 min $H_2S$ | 80 | -148 | 8.3 |
| $MoO_3$+60 min $H_2S$ | 81 | -148 | 8.1 |
| Pt | 37 | -17 | 32.7 |
| $MoS_2$ thin film [S3] | 120 | ≈ -250 | 1.5 |
| $MoO_3/MoS_2$ Core-shell [S4] | 50-60 | - 200 | ≈8 |
| $[Mo_3S_4]^{4+}$ cluster [S5] | 120 | -150 | 16.5 |
| $[Mo_3S_{13}]^{2-}$ cluster[S6] | 38-40 | -100 to -120 | ≈10 @ ≈ -200 |
| Amorphous $MoS_x$ [S7] | 60 | ≈ -150 | ≈10 @ -200 |
| Amorphous $MoS_x$ [S8] | 40 | ≈ -140 | ≈15 @ -200 |
| $MoS_2$ NPs/RGO[S9] | 41 | ≈ -100 | – |
| Mesoporous $MoS_2$[S10] | 50 | ≈ -150 | ≈ 4 @ -200 |
| Co+$MoS_2$[S11] | 101 | ≈ -300 | 4 |
| Au+$MoS_2$[S12] | 86 | -220 | ≈ 5 |
| MoP\|S[S13] | 50 | ≈ -40 | 10 @ -86 |
| $Cu_3P$ nanowires [S14] | 67 | -62 | 10 @ -143 |
| cobalt phosphosulphide [S15] | 56 | _ | 10 @ -48 |

The enhanced HER activity is further illustrated by comparing the slope of Tafel plots (Fig. 3b) and onset potentials for $H_2$ evolution reaction as shown in Table S3. The table also includes literature data of various $MoS_2$ structures. Notably, among our data the $MoO_x/MoS_2$ nanoporous electrode with 2 min-sulfurization exhibits the lowest Tafel slope of 63 mV/dec and an onset potential of -147 mV. In comparison with literature data the structure here presented performs well, particularly considering that not an extensive optimization of the sulfurization has been carried out. [S3-15] In fact, the well-conducting mixed oxide cores lead to an even more positive current onset and a higher current density, than reported for $MoS_2$-$MoO_3$ core-shell nanowires. [S4]

(a)

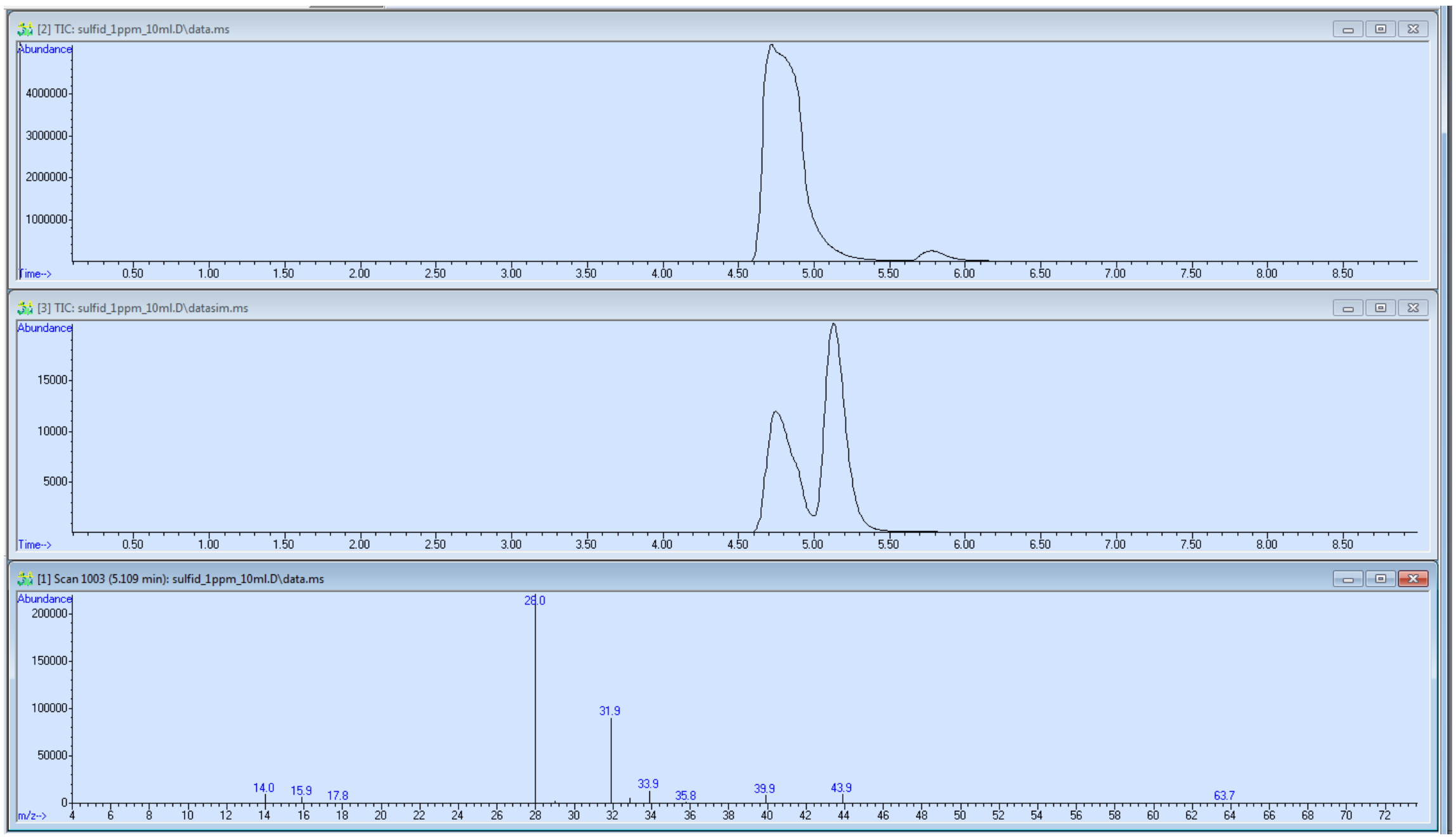

(b)

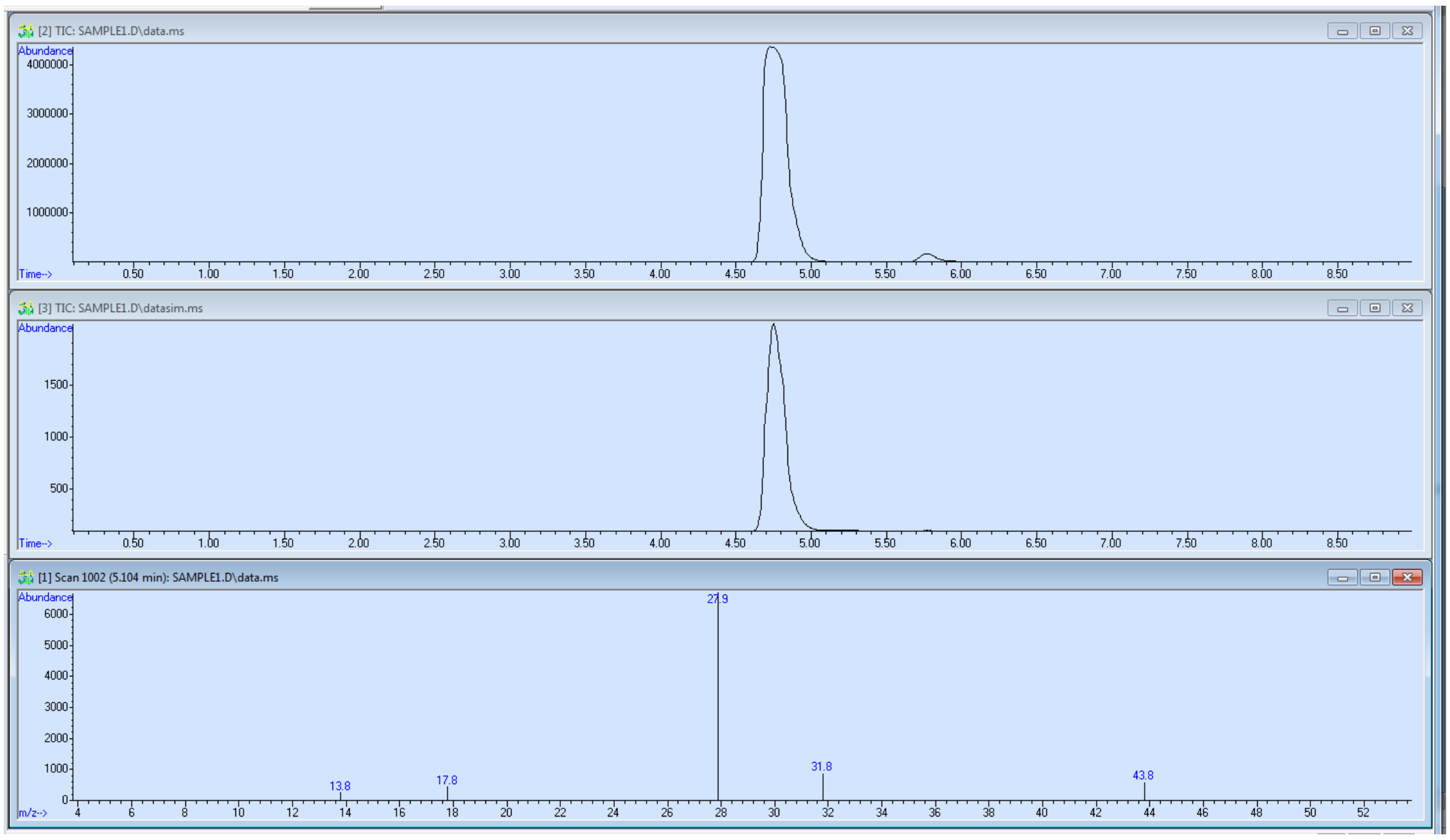

Fig. S10. GC-MS spectra (Total Ion Chromatogram, Selective Ion Monitoring and Mass Spectroscopy at retention time of 5.10 min) for (a) standard $H_2S$ gas mixture and (b) gas mixture from the head space of closed tubes with core-shell sample as working electrode after HER reactions.

Fig. S10a shows the TIC, SIM and MS spectra for standard $H_2S$ gas mixture. This standard gas mixture is obtained by reaction of $Na_2S$ and HCl, in which the concentration of $H_2S$ is low, $1.56 \times 10^{-5}$ mol/L. In the SIM spectra, there are two clear peaks at retention time of ~ 4.7 min and 5.1 min. The signal at ~ 4.7 min can be seen from standard gas and sample gas, which are ascribed to $N_2$. The difference of two samples is the signal at retention time of ~ 5.1 min. The mass spectrum at retention time of 5.10 min of standard gas shows it contains $N_2$ (m/z: 28.0), $O_2$ (m/z: 31.9), $H_2S$ (m/z: 33.9), $CO_2$ (m/z: 43.9) and $H_2O$ (m/z: 17.8) etc. For sample gas after HER reaction, in the selective ion spectrum at a retention time at 5.1 min, no clear peak can be seen. The mass spectrum at this retention time shows it contains mainly $N_2$ (m/z: 27.9), $O_2$ (m/z: 31.9), $CO_2$ (m/z: 43.9) and $H_2O$ (m/z: 17.8) etc. Accordingly, there is no $H_2S$ that can be detected from GC-MS.

The $H_2$ produced after HER reactions can be detected from GC, with a concentration 2.79 mol/L in the head space.

The presence of $H_2S$ in the gas mixture was also checked by a CV measurement, which is more sensitive. However, also with this tool, there is no $H_2S$ detected after the HER reaction. Concentration of standard solution used for CV is 10 µg/L and 20 µg/L, prepared by adding $Na_2S$• 7~8 $H_2O$ into 30% NaOH and the concentration is determined by argentometric titration.

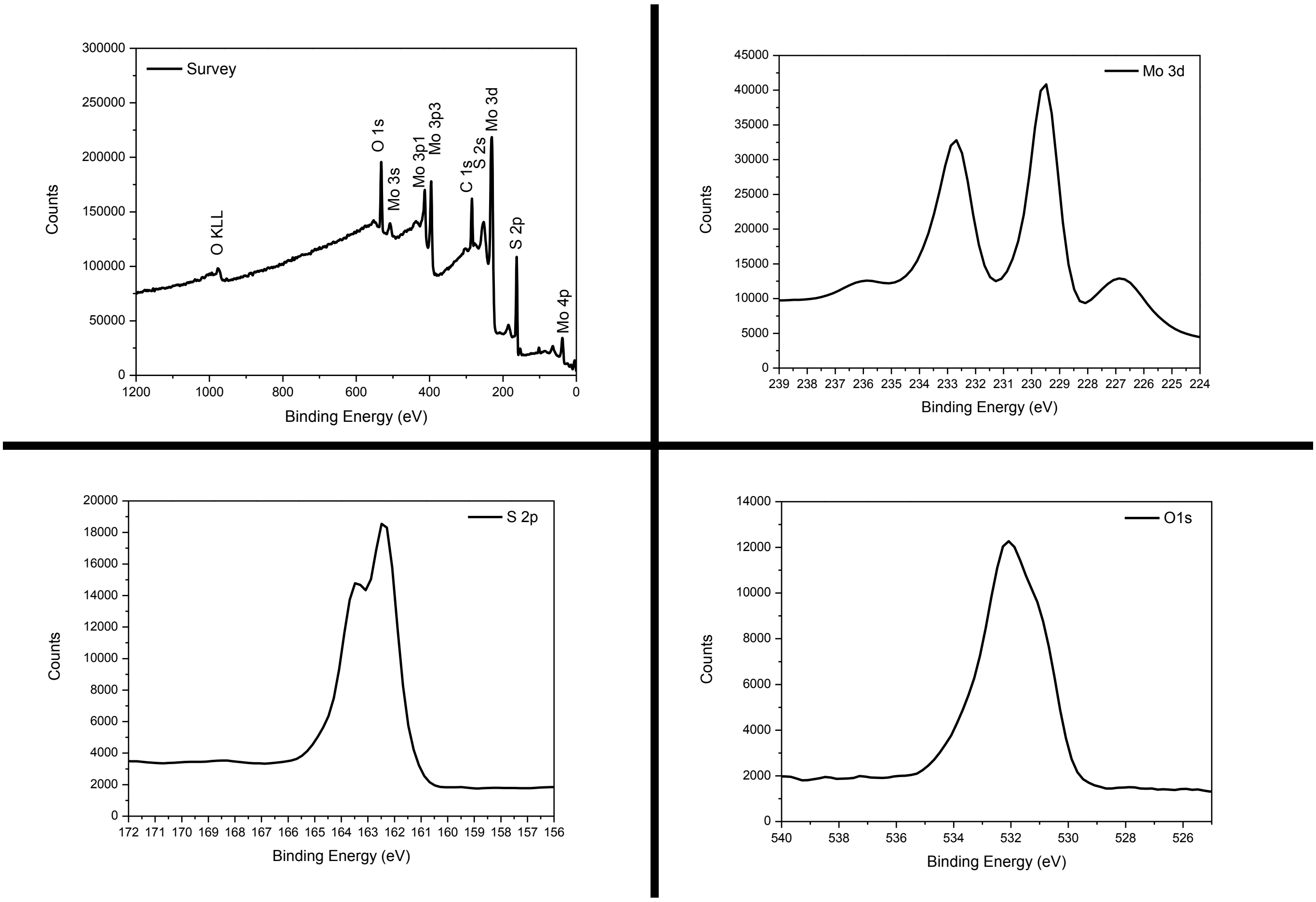

Fig. S11. XPS survey spectra (a), Mo3d (b), S2p (c), and O1s (d) peaks of core-shell MoOx/$MoS_2$ sample after HER reactions.

Fig. S11 shows that after HER reactions the chemical states of the core-shell sample remains almost the same as before. Only a mild difference could be found from the S2p peak − the peak is mainly composed of peaks at 163.8 eV and 162.5 eV, which are ascribed to Mo-S. Before the reaction, S peaks are composed of Mo-O-S and Mo-S. This indicates that during the HER reactions $MoS_2$ is more stable. Moreover, there is no significant change of the chemical state of Mo3d.

This together with the analytic results in Fig. S10 illustrates the stability of the core-shell electrode during the experiments.

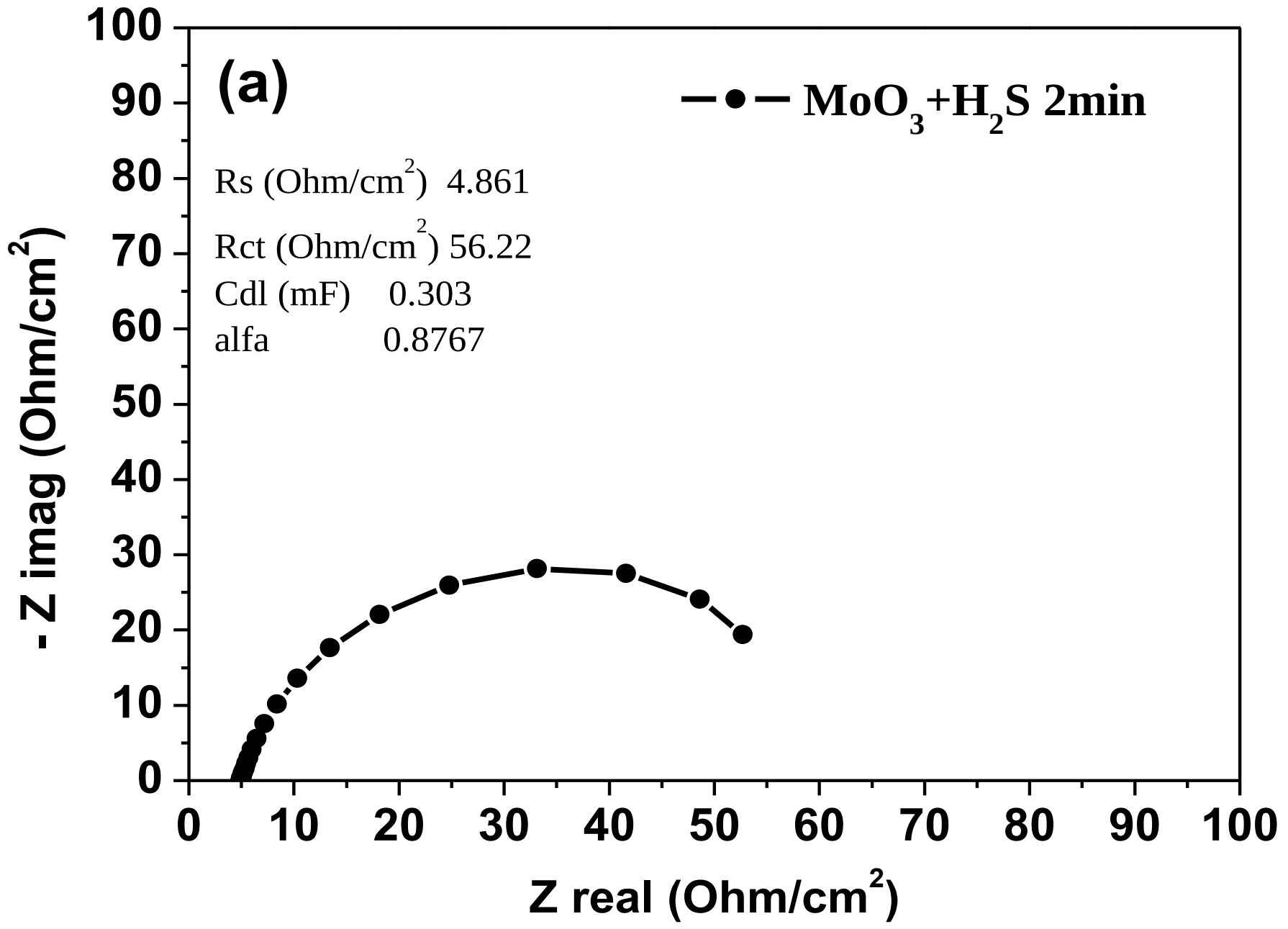

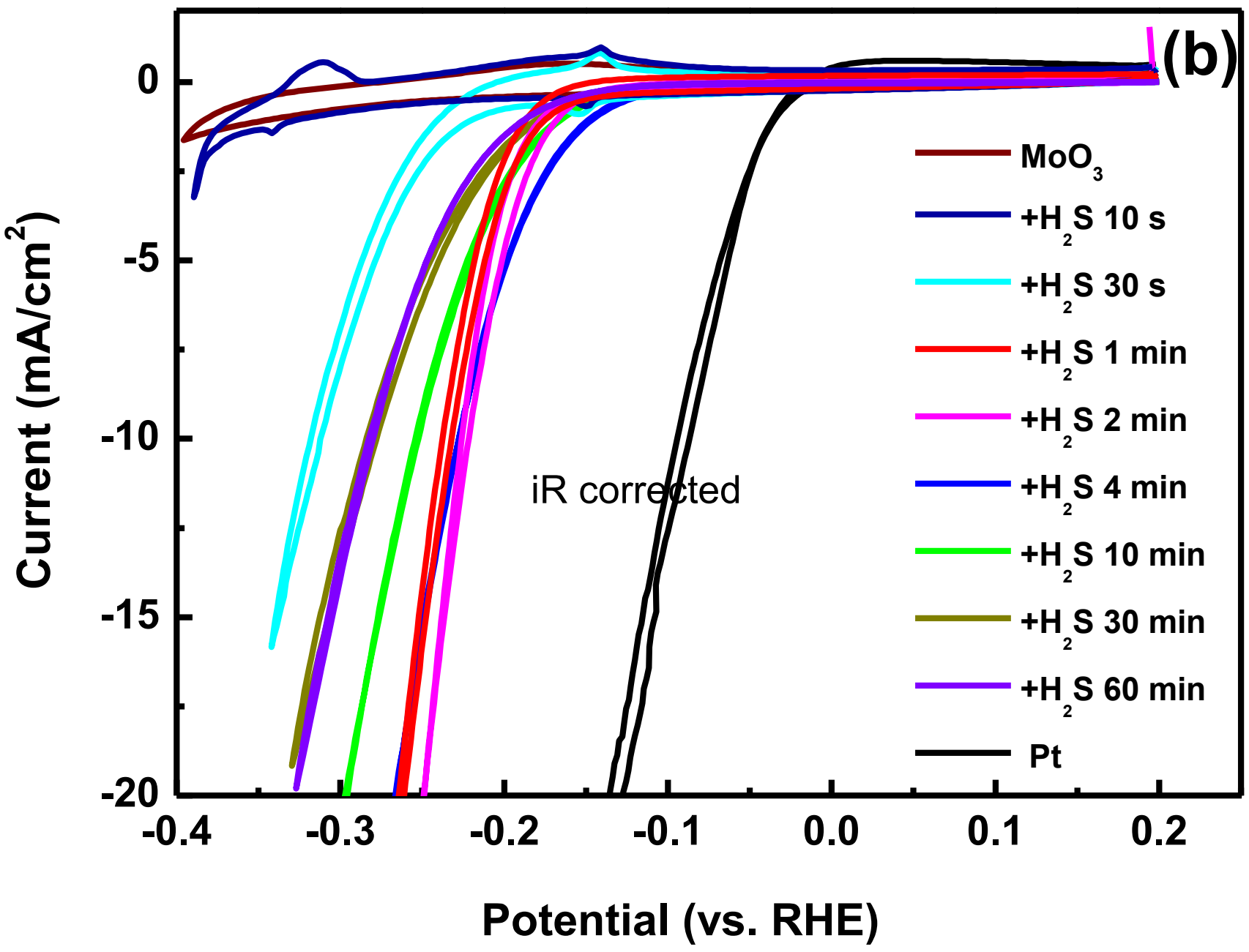

Fig. S12 (a) Impedance results for MoOx/$MoS_2$ core-shell nanoporous layer at -200 mV vs. RHE in 0.5 M aqueous $H_2SO_4$ electrolyte, inset parameters are the simulation of curve using a Randle's circuit with a constant phase element. (b) Polarization curves for self-ordered $MoO_3$ nanoporous layers under various sulfurization conditions with iR-corrected data. The solution resistance is obtained from Fig. S12a.

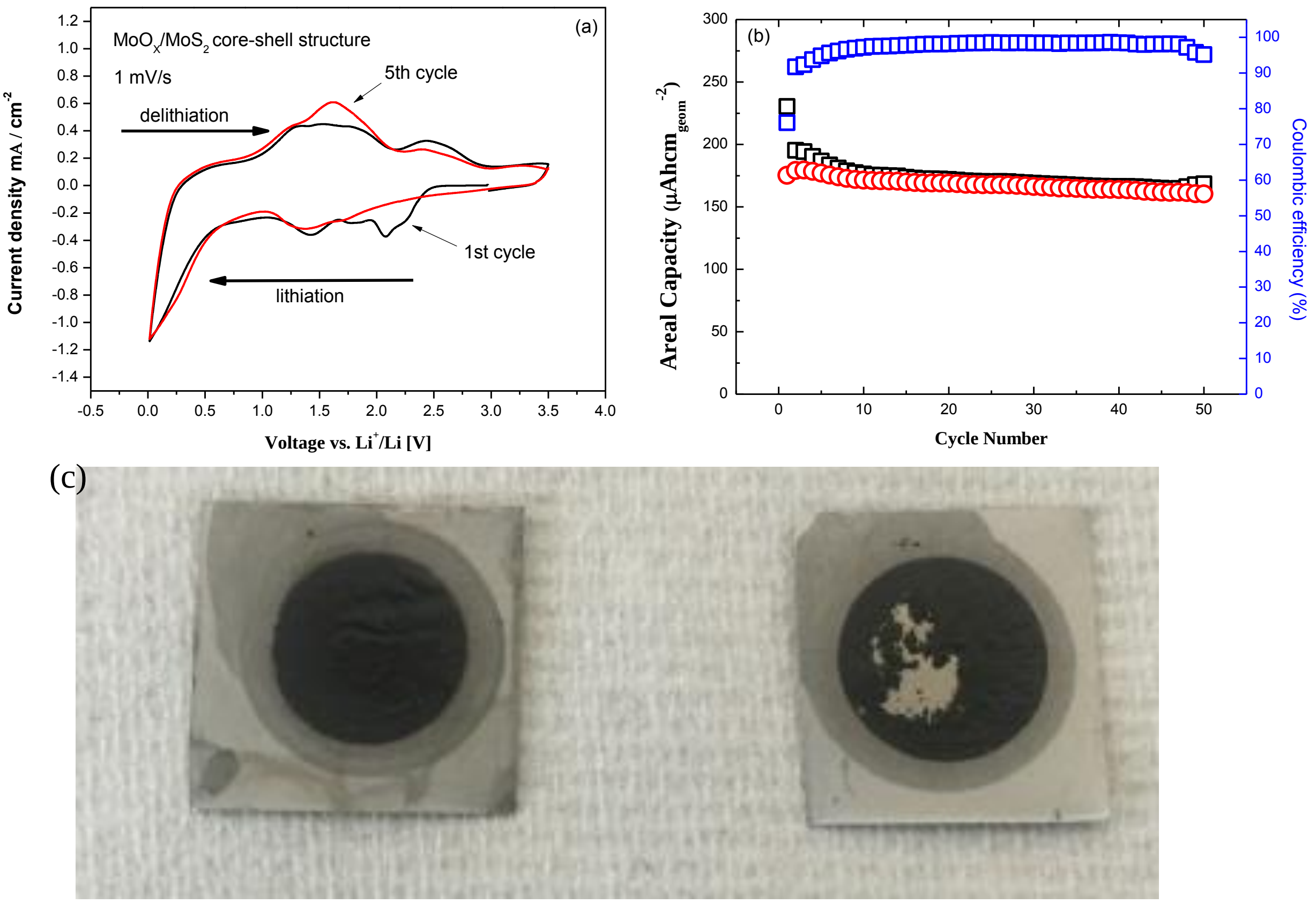

Fig. S13 (a) Cyclic voltammograms for core shell nanotube layer at a scan rate of 1 mV/s in 0.1 M $LiClO_4$ in ethylene carbonate in the window of 0.01 to 3.5 V (vs. $Li^+/Li$). (b) Charge/Discharge cycling behavior of MoOx/$MoS_2$ nanotube layer between 0.01 to 3.5 V vs. $Li/Li^+$ at a current density of 0.1 $mA/cm^2$ for 50 cycles and corresponding coulombic efficiency. (c) Optical image of nanotube layer after cycling experiments (left: MoOx/$MoS_2$ core-shell structure, Right: MoOx nanotube layer).

Fig. S13a shows the CV curves of MoOx/$MoS_2$ nanotube electrode at a scan rate of 1 mV $s^{-1}$ in the potential window of 0.01–3.5 V. During first cycle, the peaks at ~2.1 V and 1.4 V at cathodic direction can be attributed to the lithiation and reduction of $MoS_2$ and $Mo^{4+}$, respectively. In the charge process, respective oxidation and delithiation of lithiated Mo-sulfide occur as below. [S16]

$$MoS_2 + xLi^+ + xe^- = Li_xMoS_2 \qquad (1)$$

$$Li_xMoS_2 + 4Li^+ + 4e^- = 2Li_2S + Mo \qquad (2)$$

Fig. S13b shows the charge/discharge cycling behavior and corresponding coulombic efficiency of the $MoO_x$-$MoS_2$ core-shell sample. At the current density of 0.1 $mA/cm^2$, the capacity is calculated based on the geometric area of sample layer. After 50 cycles, core-shell MoOx/$MoS_2$ gives a value of 95% of coulombic efficiency. This is in line with Novak et al [S17] who has shown that a structure of $MoO_3$ coated with $MoS_2$ sheets can have a significant better cycling stability due to the thin coating – this is also clear from the optical images in the Fig. S13c.

Table **S4.** Summary of recent electrochemical data collected for Mo sulfide and oxide.[S18]

| Samples | Synthesis method | First discharge capacity (mA $hg^{-1}$) | First charge capacity (mA h $g^{-1}$) | Reversible capacity after X cycles (mA $hg^{-1}$) | Coulombic efficiency after Y cycles (%) | Current density | Highest current density tested |
|---|---|---|---|---|---|---|---|
| $MoO_x/MoS_2$ NTs[experiment] | Anodization + thermal treatment | 1156 | 880 | 846(50) | 96% (50) | 0.5 A/g | 5 A/g |
| $MoO_x/MoS_2$ NTs [estimated] | Anodization + thermal treatment | 847 | 645 | 620 (50) | 95% (50) | 0.37 A/g | 3.7 A/g |
| $MoS_2$-GNS NPs[S19] | Hydrotherm al | 2200* | 1300 | 1290(50) | 99.2 (50) | 0.1 A $g^{-1}$ | 1 A $g^{-1}$ |
| $MoS_2$-amC NPs[S16a] | solvotherma l | 1062 | 917 | 907 (50) | 87 (1) | 1.062 A $g^{-1}$ | 53.1 A $g^{-1}$ |
| $MoS_2$@CMK-3[S20] | Template - assisted | 1056 | 824 | 602 (100) | 97* (100) | 0.25 A $g^{-1}$ | 2 A $g^{-1}$ |
| a-$MoO_3$ nanobelts[S21] | hydrotherm al | 280* | 205* | 140 (100) * | 99*(100) | 0.2 A $g^{-1}$ | 0.5 A $g^{-1}$ |
| $MoO_2$[S22] | $MoO_3$ reduction | 417 | 318 | 85% | ~100% (20) | 5 $mAcm^{-2}$ | _ |
| $MoO_2$-OMC[S23] | pyrolysis | 1300* | 780* | 689 (50) | 97% (50) | 0.05 A $g^{-1}$ | 2 A $g^{-1}$ |

* is the value estimated from published graph. From these data one estimates of a charge capacity (see Fig. S13b) in the range of 620 to 846 mA h $g^{-1}$ (due to two mass assessments of the layer) at 0.1 $mA/cm^2$ (area is the geometric area of the layer) for the core shell structure used here and their comparison with literature indicate capacitance data that are very promising for further work.